\documentclass[10pt,reqno]{amsart}
\usepackage{graphicx,multicol}
\usepackage{multirow} 
\usepackage{indentfirst,csquotes}
 \usepackage{array} 

\usepackage[caption=false]{subfig}
\usepackage{amssymb,amsthm,amsmath}
\usepackage{xcolor,paralist,hyperref,fancyhdr,etoolbox}
\newtheorem{theorem}{Theorem}[]

\newtheorem{lemma}[theorem]{Lemma}

\newtheorem{remark}[theorem]{Remark}
\usepackage[round]{natbib}

\usepackage{booktabs}
\usepackage{geometry}
\usepackage{multirow}
\usepackage{makecell}
\usepackage[utf8]{inputenc}
\numberwithin{equation}{section}

\hypersetup{ colorlinks=true, linkcolor=blue, filecolor=blue, urlcolor=blue,citecolor = blue }

\usepackage{lipsum}
\usepackage{multicol}

\usepackage{listings}
\begin{document}


\title[An identifiable re-parametrization of the change-point problem]{Change-point problem:\\ Direct estimation using\\ a geometry inspired  identifiable reparameterization} 

\maketitle
\begin{center}
\author{Buddhananda Banerjee\footnotetext{corresponding author\/: bbanerjee@maths.iitkgp.ac.in}}\\
\email{bbanerjee@maths.iitkgp.ac.in}\\
\address{Department of Mathematics\\  Indian Institute of Technology Kharagpur, India-$721302$}\\

and \\

\author{Arnab Kumar Laha}\\
\email{arnab@iima.ac.in}\\
\address{Operations and Decision Sciences\\  Indian Institute of Management Ahmedabad, India-$380015$}

\end{center}

\let\thefootnote\relax

\begin{abstract}
Estimation of  mean shift in a temporally ordered sequence of random variables with a possible existence of change-point   is an important problem in many disciplines.  In the available  literature  of more than fifty years the estimation methods of  the mean shift is usually dealt as a two-step problem.  A test for the existence of a change-point  is followed by  an estimation process of the mean shift, which is known as testimator.  The problem  suffers from over parametrization.   When viewed as an estimation problem,   we establish that the maximum likelihood estimator (MLE)  always gives a false alarm indicting an existence of a change-point  in the given sequence even though there is no change-point  at all. After modelling  the parameter space  as a modified horn torus. We introduce a new method of estimation of the parameters. The newly introduced    estimation method of the mean shift  is assessed with a proper Riemannian metric on that conic manifold. It is seen that its performance is superior compared to   that  of the  MLE. The proposed method is implemented on Bitcoin data and compared its performance with the performance of the  MLE. \\

\keywords{Keywords: Change point, maximum likelihood estimator, testimator, horn torus and manifold}
\end{abstract} 

\bigskip
\setlength{\columnsep}{1.5cm} 

\section{Introduction}
\label{Introduction}
Change is an inevitable phenomenon to occur to any physical object in the universe.  The time, the amount, and the pattern of changes are of special interest to scientists to understand the dynamics of  a   system.  Sometimes, either the gradual changes are unnoticeable or abrupt changes remain undetected due to  the noise present  in the data.   A sudden change of an unknown amount  at an unknown time impacting a system  can cause significant problems Hence detection of the occurrence of change, estimation of the change-point and the changed magnitude are problems that are of significant importance and  demands a lot of attention. 
The change-point  problem has been one of the most extensively studied non-regular problems in the statistical literature. It arises in a variety of different contexts like environmental sciences \cite{jaruvskova_1997}, ecology, geology, financial markets, astronomy etc. .   The \textit{retrospective at most one change-point  (AMOC) detection problem} deals with identifying a time point (which is called the change-point ) in a given temporally-ordered sequence of independent observations, after which the distribution of the random variables undergoes an abrupt change.  In contrast, the sequential change-point  detection problem deals with the identification of a point in a sequence of independent observations, which are being gathered sequentially, after which the distribution of the random variables has undergone an abrupt change. The sequential change  point problem occurs very often in problems of surveillance and quality control. The techniques for dealing with these two different types of change-point  problems are substantially different.  In this paper, we  restrict ourselves to the retrospective AMOC set-up. 

In retrospective AMOC detection problems, the presence of a change point in the given sequence of observations needs to be affirmed along with an estimate of the point of change for further decision-making.      
As an example, consider exoplanet detection techniques that use the photometry measures of  periodic dimming of the light intensity of a star caused by a planet passing in front of the star along the line of sight from the observer.  If a sequence of independent observations of light intensity is considered the presence of a planet can be suspected if there is a sudden drop in the  light intensity or a sudden increase in the light intensity.    


In recent times, a lot of discussions have been centered on climate change and its effects. Global warming, broadly defined as an abrupt increase in the earth's average temperature, is one of the main concerns of climatologists. While the exact
cause of global warming is not known the impact of global warming depends crucially on the extent of the rise in the earth's temperature compared to the historical past. Hence, it is important to have a good estimate of the same. Statistically speaking,  not only the detection of  the time  point of change  but also the estimate of the mean shift are of vital interest. 

In the context of  the retrospective change-point detection problem different authors have enriched the literature with their proposed  procedures. \cite{chernoff_1964} estimated the current mean of a normal distribution which is subjected to changes in time.  \cite{hinkley_1970} discussed inference about the change-point in a sequence of normal random variables with fixed scale parameter. \cite{hawkins_1977} performed  the likelihood ratio test and numerical approximation to its limiting distribution. The author established that the asymptotic behavior remains the same whether the pre and post-change parameters are known or not. \cite{yao_1987} gave  an approximate distribution of the maximum likelihood estimator (m.l.e) of the change-point. It is shown that the distribution, suitably normalized, of the maximum likelihood estimator based on a large sample converges to the location of the maximum for a two-sided Wiener process when the amount of change in distribution approaches zero.  \cite{bhattacharya_1987} derived m.l.e. of a change-point in the distribution of independent random variables for general multi-parameter case. \cite{siegmund_1988} introduces a method based on the likelihood ratio statistic and  extends it to the case of independent  observations from completely specified distributions belonging to  an exponential family. Joint confidence sets for the change-point and the parameters of the exponential family are also considered. A test for change-point was also derived.
The discussion by \cite{gombay_1990}  about the maximum likelihood tests for a change in the mean of independent random
variables show  that the limit distribution is the Gumbel distribution.
\cite{horvath_1993} provided the  asymptotic distribution of the maximum likelihood ratio test statistic  to check whether the parameters of normal observations had changed at an unknown point. \cite{hartigan_1994} proposed the method of linear estimation of change-point.  His argument was  based on the limit distribution of the largest deviation between a d-dimensional Ornstein-Uhlenbeck process and the origin. 
\cite{fotopoulos_2001} carried out  change-point analysis for known parameters based on the application of Weiner-Hopf factorization identity involving the distribution of ascending and descending ladder heights, and the renewal measure in random walks.
\cite{mei_2006} considered the  problem of minimizing the frequency of  false alarms for every possible pre-change distribution when post-change distribution is specified. An asymptotically optimal procedure for one-parameter exponential families is given by the author. \cite{fotopoulos_2010} derived exact computable expressions for the asymptotic distribution of m.l.e of the change-point  when a change in the mean occurred at an unknown point of a sequence of time-ordered independent Gaussian random variables.
 For more details the reader is referred to the monographs of \cite{carlstein1994}, 
\cite{csorgo_1997} and  \cite{chen_2013}. 

The related problem of estimation of the current mean or the  mean shift has also been studied but  not to a large extent.  The estimation of mean shift  is  dependent on the estimation  of change-point, which is then considered to be a nuisance parameter.  The problem of estimating the current process mean in the presence of change-point is considered by \cite{barnard_1959} and \cite{chernoff_1964} in the context of process control. \cite{hawkins_1986} used the least square method for estimating a change in mean.
 \cite{yashchin_1995} showed that exponentially weighted moving average (EWMA) estimators are optimal in the class of linear estimators for this problem. It is also shown how EWMA estimators can be improved by other general Markovian procedures. The author discussed adaptive schemes that are based on identifying the last change-point and using the resulting last stable segment of data. A common feature of the above-mentioned methodologies is that they achieve their best efficiencies when the sample size is large and the change-point lies near the middle of the sequence of data. They usually do not perform satisfactorily when the change-point is close to either end of the data.  A similar observation is made by  \cite{bai_1994}.  The asymptotic behaviour of the likelihood ratio statistic for testing a shift in mean in a sequence of independent normal variates is studied by \cite{yao_1986}. Also, \cite{horvath_1993} reports findings  in the same context.  Many more authors  dealt with the change-point problem from the perspective of testing   with univariate \cite[see][]{page_1955,sen_1975tests,sen_1975}, multivariate  \cite[see][]{srivastava_1986,gombay_1994application,gombay_1996approximations,gombay_1996rate,jirak_2015}, time series \cite[see][]{shao_2010,aue_2013} and random field\cite[see][]{roy_2017} data. Among the recent developments the  papers by the following authors are worth mentioning  \cite{berkes_2006,aue_2009distinguishing,huvskova_2008,keshavarz_2018,killick_2012, chen_2015}.   Note that the mean-shift is equal to zero if there is no change-point in the given data. 
The reader may refer to the monograph  by \cite{wu_cusum_2007}  which gives some interesting asymptotic results on the  distribution of the estimated change-point when  the magnitude of the change is small in a linear process.

The most commonly used technique for change point detection is to first perform a test of hypothesis with the null hypothesis being that there is no change in the sequence against the alternative that there is a change point. If this test rejects the null hypothesis of no change at a specified level of significance then an estimation process (such as MLE or CUSUM) is used to estimate the change point.  Such an estimator is commonly termed as testimator since its value depends on the result of a test.  It may be noted that the value of the testimator depends on the specified level of significance of the test which may be considered as a hyperparameter.

In this paper, we propose a frequentist estimation procedure that does not involve a test of hypothesis and hence is not inherently dependent on the value of the level of significance that is imposed externally. 

When the observed sequence of independent random variables is assumed to arise from the same family of distributions then it is interesting to observe that, under some mild regularity conditions, the maximum likelihood estimator (MLE) and the  CUSUM estimator of the change point always give a false alarm about presence of a change point even though the data set does not have any.  But,  on the other hand, if the observed sequence has a change point,  it is frequently observed that the MLE performs quite well in identifying  the location of the  change point.  It is observed that the performance of an estimator, e.g.  MLE say, has the same efficiency for the true location of the change point at $r$ and $(n-r)$, where $n(>r>0)$ is the sample size and the magnitude of mean shift is non-zero.    On the other hand, the change-point locations  $r=0$ and $r=n$  both indicate the non-existence of any change-point in the data. In addition,  zero mean shift also stands for the non-existence of any change point in the data set.  The above facts show that the usual parametrization of the change point problem is non-identifiable.  In this paper,  we introduce a novel identifiable parametrization of the change point problem. We model the parameter space as a  horn-torus with a gradually evolving radius, which is a  3-dimensional  conical-manifold  \cite[see][for conical-manifold]{ghimenti_2005}.

As a consequence, the modelling becomes free from over parametrisation.  On this cone-manifold we construct  a data-driven graph with  the nodes $\{0, 1, 2,\cdots, n-1 \}$  and edges connecting $0$ with $j$ for every $j \ne 0$. These nodes correspond to  each possible location of the change point.  The structure of the above graph  will be used to define the loss function.  
We introduce an ergodic Markov chain on this graph using an associated transition probability matrix (TPM)  whose elements are derived from the values of the likelihood at different nodes.  The mode of the  stationary  distribution of this Markov chain is then used to estimate the location of the change point. The efficiency of the proposed estimator is assessed  and compared with the MLE with a suitable metric on this  cone-manifold .  When there is no change point, our method is able to identify that to a  large extent, where as the performance of the testimator is dependent on the specified level of the test. The rest of the paper is structured as follows.  Section \ref{false_alarm} discusses  how the MLE may produce a false alarm of the change-point when the data is free from it.  Section \ref{unique_parametrisation} provides the unique parametrization of the problem to remove the unwanted redundancy in the parameter space. A new estimation method based on the limiting distribution of a random walk has been introduced in Section \ref{estimation}. The  extensive simulation results are  reported  in   Section \ref{simulation}, which is followed by Bitcoin data analysis in Section \ref{data}. We have discussed how to extended our idea to dependent data in Section \ref{discussion}. 

The necessary proofs of Lemma \ref{lm_mle_ratio}, Lemma \ref{lm_cp_eatimaors}   and Theorem \ref{thm_efficiency} are provided in Appendix \ref{pf_lm_mle_ratio},\ref{pf_lm_cp_eatimaors} and \ref{pf_thm_efficiency} respectively after the concluding Section \ref{conclusion}.

\section{ A draw-back of the MLE}
\label{false_alarm}
Let $X$ be a random variable, defined on a probability space $(\Omega, \mathcal{F}, P)$, with the  cumulative distribution function (c.d.f.) $F_\mu(\cdot)$ which is absolutely continuous   with respect to the Lebesgue measure $\lambda(\mathbb{R}, \mathcal{B}(\mathbb{R}))$ providing  a density function $f_\mu(\cdot)$ on $\mathbb{R}$ equipped with the Borel $\sigma$-algebra $\mathcal{B}(\mathbb{R}).$  The model parameter $\mu$ belongs to a  parameter space $\Xi$ which is an open subset in $\mathbb{R}.$ For a sequence of time-indexed independent random variables   $X_1, X_2, \cdots, X_n$  the  change-point  problem is: 
\begin{eqnarray}
	X_1, X_2, \cdots, X_r &\stackrel{i.i.d.}{\sim}& f_{\mu_1}(\cdot)\nonumber\\
 \mbox{~~and ~~}X_{r+1}, X_{r+2}, \cdots, X_n &\stackrel{i.i.d.}{\sim}& f_{\mu_2}(\cdot),
\end{eqnarray}
where $(\mu_1, \mu_2)\in\Xi\times \Xi$ for some unknown time index  $r\in\mathbb{Z}_{n-1}=\left\{0,1,2,\cdots, (n-1)\right\}.$  This formulation involves three parameters $(\mu_1, \mu_2, r)$ which are all in general unknown. It may be noted in this parametrization, the case of no change point i.e. $r=0$ can be represented by multiple points in the parameter space  $ \Xi^2 \times \mathbb{Z}_{n-1}$ which are either of the form $(\mu_1=\mu,\mu_2=\mu,r \neq 0)$ or $(\mu_1, \mu_2,r=0)$.  This gives rise to the problem of non-identifiability in the model which makes estimation of the parameters using methods such as MLE not suitable for use in this set-up.  

For carrying out asymptotic analysis in this framework it is often assumed that when a change point is present (i.e. $r\neq 0$), then $r$ is a function of $n$  and  $\frac{r}{n}\rightarrow \nu$ as $n \rightarrow \infty$, where $0<\nu<1$.  Thus, it is assumed that as $n$ becomes large, the length of both the segments, pre and post the change-point, becomes large too.  

Lemma \ref{lm_mle_ratio}  given below points to a major drawback of the MLE when used for the change-point detection  problem. It shows that  the MLE of $r$ would lie in the set $\mathbb{Z}_{n-1} \setminus \{0\}$ with probability one. Hence, even when no change point is present in the dataset the MLE of $r$ would be different from 0  i.e. it would falsely indicate the presence of a change point, which is generally referred to as a false alarm.  


Let $X_1, X_2, \cdots, X_n$ be independent continuous random variables with density $f_{\mu_1}(\cdot)$ for $1 \le i \le k$ and $f_{\mu_2}(\cdot)$ for $k+1 \le i \le n$ where $k \in \mathbb{Z}_{n-1}$.  
Let $\lambda_0(\mathbf{x})=\displaystyle \prod_{i=1}^n f(\mu|x_i)$ be the likelihood when $k=0$ and $\lambda_k(\mathbf{x})= {\displaystyle \prod_{i=1}^{k} f(\mu_1|x_i)\prod_{i=k+1}^n f(\mu_2|x_i)}$ be that when $ { k\in  \mathbb{Z}_{n-1}\setminus\{0\}}$. 

\begin{lemma}
Assume that $\log f(\mu|x)$ is a strictly concave function of $\mu$ and has a unique maxima in the open set  $\Xi$. Let  $X_1, X_2, \cdots, X_n$ be i.i.d.  $f(\cdot|\mu)$. Then 
\begin{eqnarray}
\displaystyle\sup_{\mu\in\Xi}\lambda_0(\mathbf{x}) <  \displaystyle \max_{ k\in  \mathbb{Z}_{n-1}\setminus\{0\}}\sup_{ (\mu_1,\mu_2)\in\Xi\times \Xi } \lambda_k(\mathbf{x}) \end{eqnarray}
Thus, $P(\hat{r}_m=0)=0$ where $\hat{r}_m$ is the MLE of $r$. 
\label{lm_mle_ratio}
\end{lemma}

Proof:  Following \cite{makelainen_1981} the proof is provided in the  Appendix \ref{pf_lm_mle_ratio}.\\
  
Thus, the conventional frequentist approach to the change point problem consists of two stages. Initially, a test of hypothesis $H_0:r=0$ against $H_1: r \ne 0$ is carried out at an arbitrarily chosen level of significance ($\alpha$) and then depending on the result the second stage of estimation is executed. If $H_0$ is not rejected at $100\alpha \%$ level of significance then the data is treated to be coming from a single population and the parameter can be estimated accordingly. If $H_0$ is rejected then all three parameters $(\mu_1, \mu_2, r)$ are estimated using a suitable technique. The commonly used tests are  likelihood ratio test \cite[see][]{horvath_1993} or  the CUSUM method \cite[see][]{wu_cusum_2007}.   To the best of our knowledge, there has been no work reported in the literature that tries to deal with the change point problem purely as an estimation problem in the frequentist setup. In this paper,  we introduce a new reformulation of the parameter space of the change point problem that ensures identifiability and allows for the estimation of parameters without taking recourse to a test of hypothesis.  

\section{An identifiable parametrisation}  
\label{unique_parametrisation}
We have noted in the earlier Section \ref{false_alarm}, that the parametrization  $(\mu_1, \mu_2, r) \in \Xi^2\times  \mathbb{Z}_{n-1} $  is problematic because of non-identifiability.  Note that $(\mu_1, \mu_2, r)$  can be written as $(\mu_1, \mu_1+(\mu_2 - \mu_1), r) \equiv (\mu_1, \mu_1+\Delta, r)$. Thus we can parametrize using $(\mu_1, \Delta, r) \in \Xi\times\mathbb{R}\times \mathbb{Z}_{n-1}$. However, this does not eliminate the non-identifiability problem as the case of no change point can be represented by all points in the subset  $$\{(\mu_1,0,r): \mu_1 \in \Xi, r \in \mathbb{Z}_{n-1}\setminus\{0\}\} \cup \{(\mu_1,\Delta,0): \mu_1 \in \Xi, \Delta \in \mathbb{R} \}$$
Now, observe that in the context of the change point problem, the parameters $\Delta$ (magnitude of change) and $r$ (location of change point) are the parameters of our interest whereas the parameter $\mu_1$ is a nuisance parameter. 

We now introduce a different parametrization that overcomes the problem of non-identifiability.  Let $\tau$  be the left closed right open interval $[0,1)$.   Note  that  an equivalent way of representing a change point at $r$ is to write it as the ratio $\frac{r}{n} \in \tau$.  



Let  $\Theta$ be the open interval $(-\frac{\pi}{2}, \frac{\pi}{2})$ and define $\theta =\arctan(\Delta) \in \Theta.$ To identify the points in $\tau\times \Theta$  which represent the ``no-change" situations define a mapping  $\mathbf{U}: ~\tau\times \Theta \mapsto \mathcal{M} \subset \mathbb{R}^3$ as, 
 $\mathbf{U }(t,\theta)= c(t)\mathbb{D}(\mathbf{m}(\theta)) e(t)$ 
 where, $c(t)=t(1-t)$,   $\mathbb{D}(\mathbf{m}(\theta))$ is a diagonal matrix with diagonal vector $\mathbf{m}(\theta)=(1-\cos\theta, 1-\cos\theta, \sin\theta)$ and, $e(t)=[{\cos(2\pi t), \sin(2\pi t), 1}]^T$. All the points but $\mathbf{0}$ in $\mathcal{M}$  indicates the exact location of the change-point. As explained below $\mathbf{0}=(0,0,0)^T$ uniquely identifies the no-change case. 
 
 The term $c(t)=t(1-t)$,  measures the closeness of the change point to the middle of the sequence. The diagonal vector 
 \begin{eqnarray}
    \mathbf{m}(\theta)&=&(1-\cos\theta, 1-\cos\theta, \sin\theta) \nonumber\\
    &=&\left(1-\frac{1}{\sqrt{1+\Delta^2}},1-\frac{1}{\sqrt{1+\Delta^2}}, \frac{2\Delta}{\sqrt{1+\Delta^2}} \right) \nonumber
 \end{eqnarray}
 captures the amount of change with the first two components depending only on the magnitude of the change (i.e. $|\Delta|$) and the third component indicates the sign of the change. Finally, $e(t)=[{\cos(2\pi t), \sin(2\pi t), 1}]^T$  maps $\tau$ into a unit circle in $(x,y,1)$ plane and the exact location of the change point is mapped to a point on this unit circle.  So, $\mathbf{U }(t,\theta)\equiv[u_1, u_2,u_3]^T\equiv\mathbf{u},$ can be represented in  matrix notation as 
 \begin{small}
\begin{equation}
\begin{bmatrix}
u_1(t,\theta)\\ u_2(t,\theta) \\ u_3 (t,\theta)
\end{bmatrix}=t(1-t)\begin{bmatrix}
1-\cos \theta & 0 &0 \\
0 & 1-\cos \theta  &0 \\
0 & 0 & \sin \theta
\end{bmatrix}   \begin{bmatrix}
\cos (2\pi t) \\\sin (2\pi t)\\ 1 
\end{bmatrix} = t(1-t) \begin{bmatrix}
(1-\cos\theta)\cos (2\pi t) \\ (1-\cos\theta)\sin (2\pi t)\\ \sin\theta
\end{bmatrix}   \label{horntorus}
\end{equation}
\end{small}

\begin{figure}[t!]
\centering
\includegraphics[height=6cm,width=6cm]{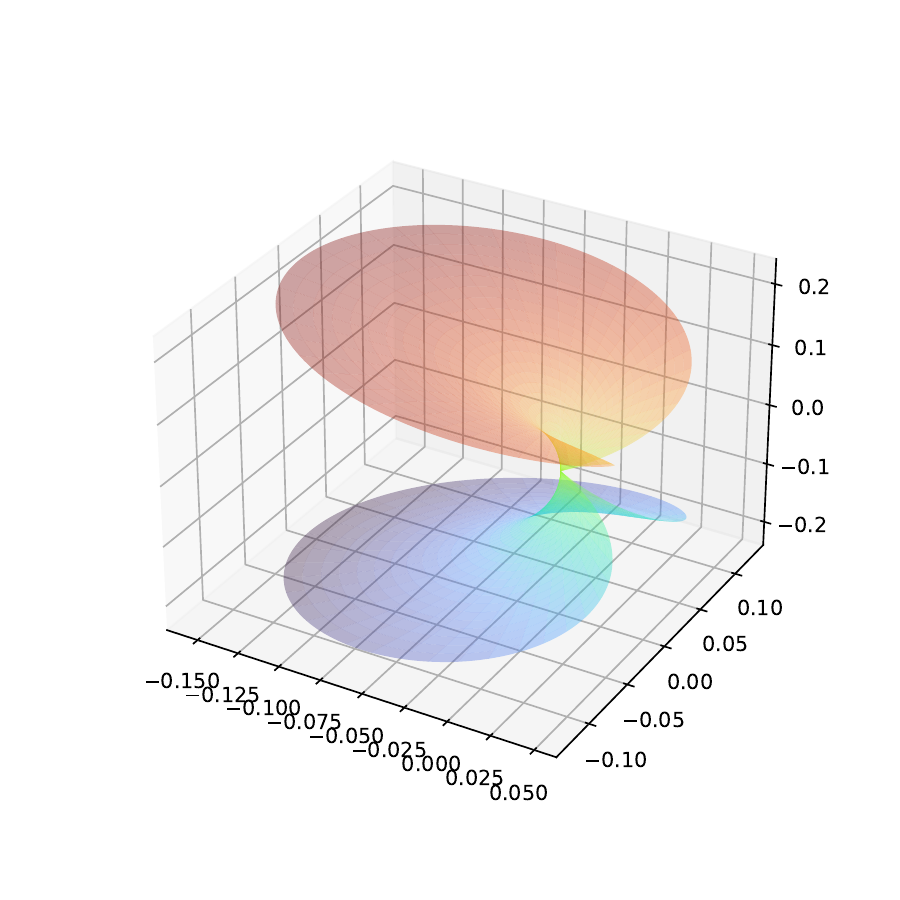} 
\includegraphics[height=6.5cm,width=6.5cm]{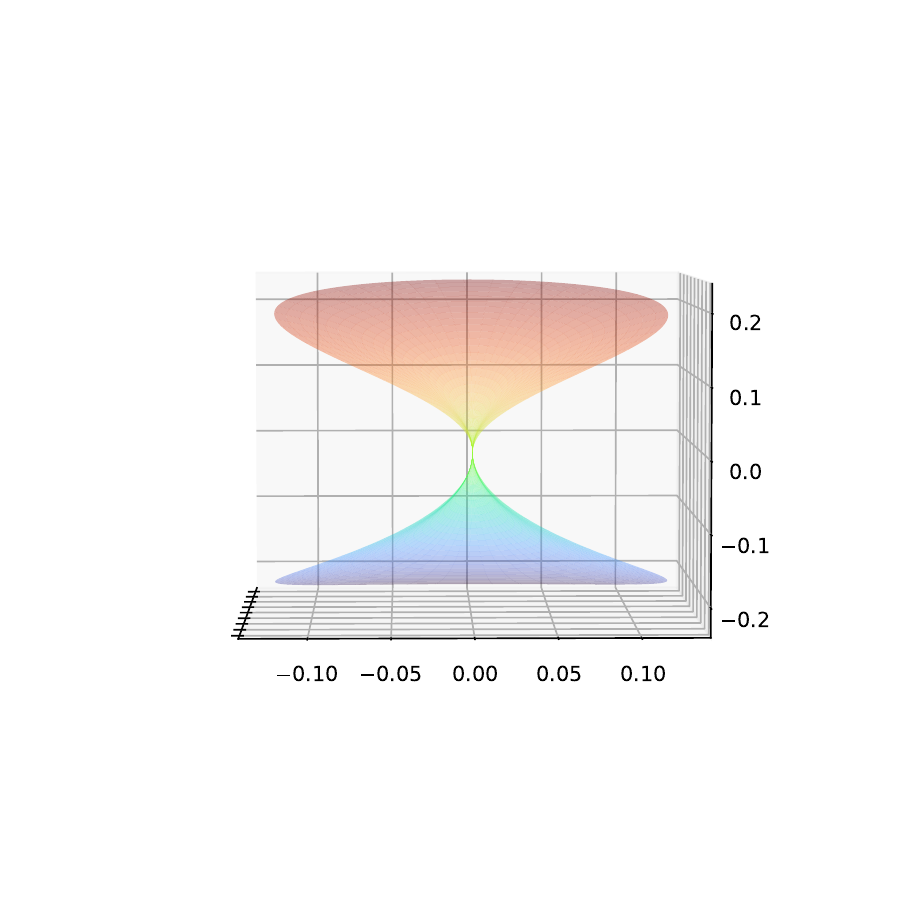}
\caption{ The parameter space $\mathcal{M}$ using Equation-\ref{horntorus}. [from different angles] }
\label{fig:paramspace}
\end{figure}

It may be noted that $\mathcal{M} \subset \mathbb{R}^3$  evolves from the equation of a horn torus\\ $[(1-\cos\theta)\cos (2\pi t), (1-\cos\theta)\sin (2\pi t), \sin\theta]^T$  but  with a changing radius $c(t)=t(1-t)$.  Note that the no-change situation  represented  by  $t=0$ and/or $\theta=\arctan(\Delta)=0$ is identified to the point $\mathbf{0}=(0,0,0)^T\in \mathcal{M}.$ But any other point on $\mathcal{M}$ uniquely describes a certain amount of change corresponding to $\theta \in \Theta$ at a particular location $t\in(0,1)=\tau \setminus \{0\}$. The parameter space can be visualized as a petal-like shape along with its reflection as shown  in    Figure \ref{fig:paramspace}.  Now, in the opposite direction,  given  $\mathbf{u}=(u_1, u_2, u_3)^T\neq\mathbf{0} \in \mathcal{M}$  we can obtain  $(\theta,t)$  or equivalently $(\Delta, r)$ as follows: $$\tan(\theta)=\Delta=\frac{u_3}{t(1-t)-\sqrt{u_1^2+u_2^2}} \mbox{~where~} t=\frac{r}{n}=\frac{1}{2\pi}\arctan\left(\frac{u_2}{u_1}\right).$$  Because of the point of singularity, $\mathbf{0}$, the surface $\mathcal{M} $ is not a smooth manifold. But it is a special case of a conical manifold, for the definition  of which and  other properties  the reader may look at  \cite{ghimenti_2005}. 


\section{ Estimation and main result }
\label{estimation}

 For a sequence of time-indexed independent random variables   $X_1, X_2, \cdots, X_n$ consider the  at-most-one change-point  problem as discussed in  Section \ref{false_alarm}.
The likelihood function for this problem is   
\begin{eqnarray}
  \begin{cases}
\ell(0)=\displaystyle \max_{\mu \in \Xi } \prod_{i=1}^n f(\mu|x_i) \mbox{~~if~~} \mathbf{u}=\mathbf{0}, \\
\ell(k)=\displaystyle \max_{\mu_1 \in \Xi } \prod_{i=1}^k f(\mu_1|x_i)\max_{\mu_2 \in \Xi } \prod_{i=k+1}^n f(\mu_2|x_i)\mbox{~~if~~} \mathbf{u}\neq\mathbf{0} \mbox{~i.e.~} t=\frac{k}{n}\neq0, \mu_2-\mu_1\neq 0
\end{cases}
\end{eqnarray}
where, $k\in \{1,2, \ldots,(n-1)\} $ and $ \mathbf{u}$ is defined in Equation \ref{horntorus}.   According to the Lemma \ref{lm_mle_ratio}, the  index of the  maximum of the  likelihood sequence $\boldsymbol{\ell}=(\ell(0), \ell(1), \cdots, \ell(n-1)) $, always  lies in  $ \mathbb{Z}_{n-1} \setminus \{0\}$, even though there is no change-point  in the data.  Hence, the  MLE of the change-point  is 
\begin{equation}
	\hat{r}_m=  \arg\max_{k \in \mathbb{Z}_{n-1} } \ell(k)  \label{rmle}
\end{equation}
The corresponding estimator of the mean shift is 
\begin{equation}
	\hat{\Delta}_{\hat{r}_m}=  \displaystyle	\arg \max_{\mu_2 \in \Xi } \prod_{i={\hat{r}_m}+1}^n f(\mu_2|x_i)- \displaystyle	\arg\max_{\mu_1 \in \Xi } \prod_{i=1}^{\hat{r}_m}f(\mu_1|x_i)\label{delmle}
\end{equation}
As a consequence of the  Lemma \ref{lm_mle_ratio}, $\hat{\Delta}_{\hat{r}_m}\neq 0$ and $\hat{\mathbf{u}}_m\neq \mathbf{0}$ with probability one. To see this, note that $\hat{t}_m=\frac{\hat{r}_m}{n}\in\tau \setminus \{0\}$ and $\hat{\theta}_m=\arctan(\hat{\Delta}_{\hat{r}_m} )\in (-\frac{\pi}{2}, \frac{\pi}{2})$. Hence, the estimated parameter point  $\hat{\mathbf{u}}_m=\mathbf{U}(\hat{t}_m,\hat{\theta}_m) \in \mathcal{M} \setminus \{\mathbf{0}\}.$

To overcome this shortcoming  let us consider a graph $\mathcal{G}$ on $\mathbb{Z}_{n-1}$  with a symmetric adjacency matrix $A_{n\times n}$ defined as 
\begin{equation}
	A= ((a_{ij}))= \begin{cases}
	1 \mbox{~~if~~} |i-j|= 0 \mbox{~~or~~} i\cdot j=0,\\
	0 \mbox{~~ otherwise.~~}
	\end{cases}
 \label{adj_matrix}
\end{equation}
Hence, all nodes of  $\mathcal{G}$ are not only self-connected but each node is connected to  $\{0\}$ also. Now to construct a random walk  on  $\mathcal{G}$, let us first define a transition probability matrix $\mathcal{T}$ depending on both the adjacency matrix $A$ and the likelihood sequence $\boldsymbol{\ell}$ as $	\mathcal{T}= [ \mathbb{D}(A\boldsymbol{\ell})]^{-1} [A\mathbb{D}(\boldsymbol{\ell})] $  where  $\mathbb{D}(\boldsymbol{\ell})$  and $ \mathbb{D}(A\boldsymbol{\ell})$ are diagonal matrices with diagonal vectors $\boldsymbol{\ell}$ and  $A\boldsymbol{\ell}$ respectively. The definition of $\mathcal{T}$ ensures that each of its row sum equals one.  Note that the  Markov chain  with the transition probability matrix $\mathcal{T}$ is aperiodic, irreducible on a finite state space $\mathbb{Z}_{n-1}$,  and hence   is ergodic.  So, there exists a stationary  distribution $\boldsymbol{\pi}=(\pi(0), \pi(1), \cdots, \pi(n-1))^{T}$ satisfying  $\boldsymbol{\pi}=\boldsymbol{\pi}\mathcal{T}.$
 We now obtain the value of $\pi(k)$ explicitly. For this, denote 
 \begin{equation}
 \displaystyle S=\sum_{i=0}^{n-1}\ell(i)~~ \mbox{and} ~~S_i= \ell(0)+\ell(i) \mbox{~~for~~} i=1,2,\dots, (n-1). 
 \label{likelihood identity}
 \end{equation}
Since  $\boldsymbol{\pi}=\boldsymbol{\pi}\mathcal{T}$ we get 
\begin{eqnarray*}
  \pi(i)&=&\pi(0)\frac{\ell(i)}{S}+ \pi(i)\frac{\ell(i)}{S_i}\\
  \implies \pi(i)&=&\frac{S_i}{S}\frac{\ell(i)}{\ell(0)} \pi(0)
\end{eqnarray*}
along with the identity $\displaystyle\sum_{i=0}^{n-1}\pi(i)=1.$ Solving we get 
\begin{eqnarray}
    \pi(0)&=& \displaystyle\left(1+ \sum_{i=1}^{n-1}\frac{S_i}{S}\frac{\ell(i)}{\ell(0)}\right)^{-1}   \nonumber\\
    &=& S  \ell(0)  \left(\sum_{i=0}^{n-1}\ell(i)^2+2\ell(0)(S-\ell(0))\right)^{-1} \mbox{~using identities from Equation-\ref{likelihood identity}} \nonumber\\
    &=& L(0)  \left(\sum_{i=0}^{n-1}L(i)^2+2 L(0)(1-L(0))\right)^{-1} \\
\mbox{~and~} \pi(i)&=& \frac{\left(L(i)^2+L(0)L(i) \right) }{\displaystyle\sum_{i=0}^{n-1}L(i)^2+2 L(0)(1-L(0))}\mbox{~~ for~} i=1,2,\ldots,(n-1).
\label{pi_vector} 
\end{eqnarray}
with $L(i)=\ell(i)/S$ for $i=0,1,2,\ldots,(n-1).$

Now, based on this stationary distribution $\boldsymbol{\pi}$ we propose  new estimators for   $r$  and  $\Delta$ respectively as 
\begin{equation} 
	\hat{r}= \arg\max_{k \in \mathbb{Z}_{n-1} } \pi(k)  \label{newrhat}
\end{equation} 
 and 
 \begin{equation}
 \hat{\Delta}= \begin{cases}
 	0 \hspace{1.3cm} \mbox{~~if~~}\hat{r}=0, \\
 \displaystyle	\hat{\Delta}_{\hat{r}_m} \hspace{1.0cm} \mbox{~~if~~} \hat{r} = \hat{r}_m\in \mathbb{Z}\setminus \{0\}
 	\end{cases}
 \end{equation}
 As a consequence, we have  $\hat{t}=\frac{\hat{r}}{n}\in\tau$ and $\hat{\theta}=\arctan\hat{\Delta} \in (-\frac{\pi}{2}, \frac{\pi}{2})$. Hence the estimated parameter point on $\mathcal{M}$ is $\hat{\mathbf{u}}=\mathbf{U}(\hat{t},\hat{\theta}).$The next  Lemma \ref{lm_cp_eatimaors} establishes a connection between  the MLE, $\hat r_m$, and the proposed estimator $\hat r$.


\begin{lemma}
$\hat{r}=0$ or $\hat{r}_m$ with probability 1. 
\label{lm_cp_eatimaors}
\end{lemma}
 Proof: The proof of the lemma is  based on a key finding that $\{L(1),L(2), \ldots, L(n-1)\}$ and  $\{\pi(1),\pi(2), \ldots ,\pi(n-1)\}$ have the same ordering with respect to their magnitudes leading to the only options $\hat r= 0$ or $\hat r_m .$ 
The details of the proof are  provided in Appendix \ref{pf_lm_cp_eatimaors}.

Let $\mathbf{p}_{L}$ and $\mathbf{p}_{\pi}$ be the probability vectors representing the marginal distributions of  $\hat{r}_m$ and  $\hat{r}$ for the location of the change point. From  Lemma \ref{lm_mle_ratio} we know  that $\mathbf{p}_{L}(0)=P(\hat{r}_m=0)=0,$ whereas, in general,  $\mathbf{p}_{\pi}(0)=P(\hat{r}=0)\geq 0.$ 


\begin{lemma}
$P(\hat{r}=0)= \displaystyle \sum_{k=1}^{n-1}\delta_k~ p_L(k) $  where $\delta_k=P\left(\hat{r}=0|\hat{r}_m=k\right).$
\label{lm_r_zero}
\end{lemma}
Proof: Define the conditional probability , $P(\hat{r}=0|\hat{r}_m=k)=\delta_k$ and  observe that 
\begin{eqnarray}
    P\left(\hat{r}=0\right)&=& P\left(\{\hat{r}=0\},\displaystyle \bigcup_{k=1}^{n-1}\{\hat{r}_m=k\}\right)= P\left(\displaystyle \bigcup_{k=1}^{n-1}\{\hat{r}=0, \hat{r}_m=k\}\right)\nonumber\\
    &=&\displaystyle \sum_{k=1}^{n-1}P\left(\hat{r}=0|\hat{r}_m=k\right)P\left(\hat{r}_m=k\right)\nonumber\\
    &=&\displaystyle \sum_{k=1}^{n-1}\delta_k~ p_L(k)  
    \label{zero_prob}
\end{eqnarray}

For ${\mathbf{u}_1}=\mathbf{U}({t}_1,{\theta}_1)$ and ${\mathbf{u}_2}=\mathbf{U}({t}_2,{\theta}_2)$ both on $\mathcal{M}$  let us define a metric, 
\begin{equation}
	d(\mathbf{u}_1, \mathbf{u}_2)= \begin{cases} 
	t(1-t)|\theta_1-\theta_2|  \hspace{2cm}\mbox{~~if~~}  t_1=t_2=t,\\ 
	t_1(1-t_1)|\theta_1|+ t_2(1-t_2)|\theta_2|  \hspace{0.2cm} \mbox{~otherwise,}
	\label{zero_pass_metric}
	\end{cases}
\end{equation}
where the distance between $\mathbf{u}_1$ and $\mathbf{0}$ is measured by  $t_1(1-t_1)|\theta_1|$ and similarly for $\mathbf{u}_2.$  It is straightforward to see that $(\mathcal{M},d)$ is a metric space. 
When $t_1\neq t_2$  or $\theta_1\theta_2<0$  then two points are  always connected  through $\mathbf{0}=(0,0,0).$

Hence this metric is a natural one for the adjacency  matrix (see, Equation  \ref{adj_matrix}) of  the embedded graph for the random walk. We will call the metric $d(\cdot,\cdot)$, defined in Equation \ref{zero_pass_metric} as  \textit{zero-pass-metric} on $\mathcal{M}$. Finally, the accuracy  of the MLE $\hat{\mathbf{u}}_m=\mathbf{U}(\hat{t}_m,\hat{\theta}_m)$  and that of the proposed estimator  $\hat{\mathbf{u}}=\mathbf{U}(\hat{t},\hat{\theta})$ are comparable on the parameter space $\mathcal{M}$ with respect to \textit{zero-pass-metric}  $d(\cdot,\cdot)$.

Let us introduce two  indicator functions  
\begin{equation}
    \mathbf{I}_k=\begin{cases}
        1 \mbox{~~~~if~~} \hat{r}_m=k  \\
        0 \mbox{~~~~~ otherwise~~}
    \end{cases}
\end{equation}
 and 
 \begin{equation}
    \mathbf{I}_{k0}=\begin{cases}
        1 \mbox{~~~~if~~} \hat{r}=0 \mbox{~when~} \hat{r}_m=k  \\
        0 \mbox{~~~~~ otherwise.~~}
    \end{cases}
\end{equation}
Further,  define,  $\mathbf{I}_{kk}=1-\mathbf{I}_{k0}$, which is the indicator of the event $\hat{r}=k$ when $\hat{r}_m=k.$ Let the true value of the parameter  $\mathbf{u}=\mathbf{U}(\frac{r}{n},\tan^{-1}\Delta) \in $  $\mathcal{M}$. 
Now, the loss function for the MLE would be defined as 
\begin{eqnarray}
    d(\hat{\mathbf{u}}_m, \mathbf{u})=&\displaystyle\sum_{k=1,k\neq r}^{n-1}&\left\{\left(\frac{k}{n}\right)\left(1-\frac{k}{n}\right)|\tan^{-1}\hat{\Delta}_k| + \left(\frac{r}{n}\right)\left(1-\frac{r}{n}\right)|\tan^{-1}{\Delta}|\right\}\mathbf{I}_k \nonumber\\
    &+&\left(\frac{r}{n}\right)\left(1-\frac{r}{n}\right)| \tan^{-1}\hat{\Delta}_r-\tan^{-1}{\Delta}|\mathbf{I}_{r}
    \label{loss_rm}
\end{eqnarray}
and similarly the loss function of the proposed estimator is 
\begin{eqnarray}
    d(\hat{\mathbf{u}}, \mathbf{u})=&\displaystyle\sum_{k=1,k\neq r}^{n-1}&\left\{\left(\frac{k}{n}\right)\left(1-\frac{k}{n}\right)|\tan^{-1}\hat{\Delta}_k|\mathbf{I}_{kk} + \left(\frac{r}{n}\right)\left(1-\frac{r}{n}\right)|\tan^{-1}{\Delta}|\right\}\mathbf{I}_{k} \nonumber\\
    &+&\left(\frac{r}{n}\right)\left(1-\frac{r}{n}\right)\left\{| \tan^{-1}\hat{\Delta}_r-\tan^{-1}{\Delta}|\mathbf{I}_{rr}+|\tan^{-1}{\Delta}|\mathbf{I}_{r0}\right\}\mathbf{I}_{r}.
    \label{loss_r}
\end{eqnarray}
Now we compare the risks of $\hat{\mathbf{u}}_m$ and $\hat{\mathbf{u}}$ as
\begin{eqnarray}
   & & E(d(\hat{\mathbf{u}}_m, \mathbf{u})-d(\hat{\mathbf{u}}, \mathbf{u}))\nonumber\\
   &=&\displaystyle\sum_{k=1,k\neq r}^{n-1}E\left[\left(\frac{k}{n}\right)\left(1-\frac{k}{n}\right)|\tan^{-1}\hat{\Delta}_k| \mathbf{I}_{k} \mathbf{I}_{k0} \right] \nonumber\\
    &+&\left(\frac{r}{n}\right)\left(1-\frac{r}{n}\right) E \left[ \left\{| \tan^{-1}\hat{\Delta}_r-\tan^{-1}{\Delta}|-|\tan^{-1}{\Delta}|\right\}\mathbf{I}_{r} \mathbf{I}_{r0} \right] \nonumber\\
    &=&\displaystyle\sum_{k=1,k\neq r}^{n-1}\left[\left(\frac{k}{n}\right)\left(1-\frac{k}{n}\right)E|\tan^{-1}\hat{\Delta}_k| p_L(k)\delta_k \right]\nonumber\\
    &+&\left(\frac{r}{n}\right)\left(1-\frac{r}{n}\right) \left[ \left\{E| \tan^{-1}\hat{\Delta}_r-\tan^{-1}{\Delta}|-|\tan^{-1}{\Delta}|\right\}p_L(r)\delta_r \right]\label{risk_difference}
\end{eqnarray}

We have $$E\left[\left(\frac{k}{n}\right)\left(1-\frac{k}{n}\right)|\tan^{-1}\hat{\Delta}_k| \mathbf{I}_{k} \mathbf{I}_{k0} \right]=\left(\frac{k}{n}\right)\left(1-\frac{k}{n}\right) E|\tan^{-1}\hat{\Delta}_k| p_L(k)\delta_k$$ because the conditional distribution of  $(\hat{\Delta}_{\hat{r}_m}|\hat{r}_m=k)$ is same as the unconditional  distribution $\hat{\Delta}_k$ where, 
\begin{equation}
	\hat{\Delta}_{k}=  \displaystyle	\arg \max_{\mu_2 \in \Xi } \prod_{i ={k}+1}^n f(\mu_2|x_i)- \displaystyle	\arg\max_{\mu_1 \in \Xi } \prod_{i=1}^{k}f(\mu_1|x_i)
\end{equation}
and $E(\mathbf{I}_{k} \mathbf{I}_{k0})=p_L(k)\delta_k$ for all $k\in \{1,2,\ldots , (n-1)\}.$

In particular, suppose $X_1, X_2, \cdots, X_n$ is an independent Gaussian sequence of random variables with known variance $\sigma^2$ assumed to be 1.  For some $r\in\mathbb{Z}_{n-1}$ (or equivalently $\frac{r}{n}=t\in \tau$),  assume that $X_1, X_2, \cdots, X_r$ are independently distributed $N(\mu_1,1)$ and $X_{r+1}, X_{r+2}, \cdots, X_n$ are  $N(\mu_2,1)$ random variables.  The amount of change or mean shift  is $\Delta=\mu_2-\mu_1=\tan \theta  \in \mathbb{R}$ for some $\theta\in \Theta.$ If $\sigma^2\neq1$ then we measure the mean shift $\Delta$ in standard deviation unit i.e. $\Delta=(\mu_2-\mu_1)/\sigma.$

Let $\eta_n(r)=\left(\frac{r}{n}\right)\left(1-\frac{r}{n}\right)\left\{E| \tan^{-1}\hat{\Delta}_r-\tan^{-1}{\Delta}|-(|\tan^{-1}{\Delta}|+E|\tan^{-1}\hat{\Delta}_r|)\right\}p_L(r)\delta_r.$ We have the following theorem:

\begin{theorem}
 If $P(\hat{r}=0)\neq 0$ and   $d(\cdot,\cdot)$ is the zero-pass-metric defined on the parameter space  $\mathcal{M}$, then   $E(d(\hat{\mathbf{u}}_m, \mathbf{u})-d(\hat{\mathbf{u}}, \mathbf{u})) \ge \eta_{n}(r)$ where $\eta_n(r) \rightarrow 0$ as $n \rightarrow \infty$.

\label{thm_efficiency}
\end{theorem}
Proof: See Appendix \ref{pf_thm_efficiency} for the proof.  Also note that, if  $P(\hat{r}=0)=0$ then  from the Lemma \ref{lm_cp_eatimaors} we have $P(\hat{r}=\hat{r}_m)=1$  which implies $E(d(\hat{\mathbf{u}}_m, \mathbf{u}))= E(d(\hat{\mathbf{u}}, \mathbf{u}))$  

\begin{remark}
The method of estimating the change-point  using  the mode of the stationary distribution of a Markov chain with transition probability matrix dependent on the likelihood has not been explored before in the literature to the best of our knowledge. The results discussed in Section \ref{Empirical_Study } shows that this is promising alternative to the MLE. We conjecture that this  technique may be useful in other non-regular problems involving  both discrete and continuous  parameters. 
\end{remark}

\section{Empirical  Study }
\label{Empirical_Study }
In this section we report the summary  of an extensive simulation study and data analysis. We perform   comparative studies of the performances of the MLE  and the proposed estimator of the change point for different parameter specifications and sample sizes. We also extend  our idea to the situation when population variance is fixed but unknown.

\begin{figure}[t!]
\centering
\includegraphics[height=8cm,width=7 cm]{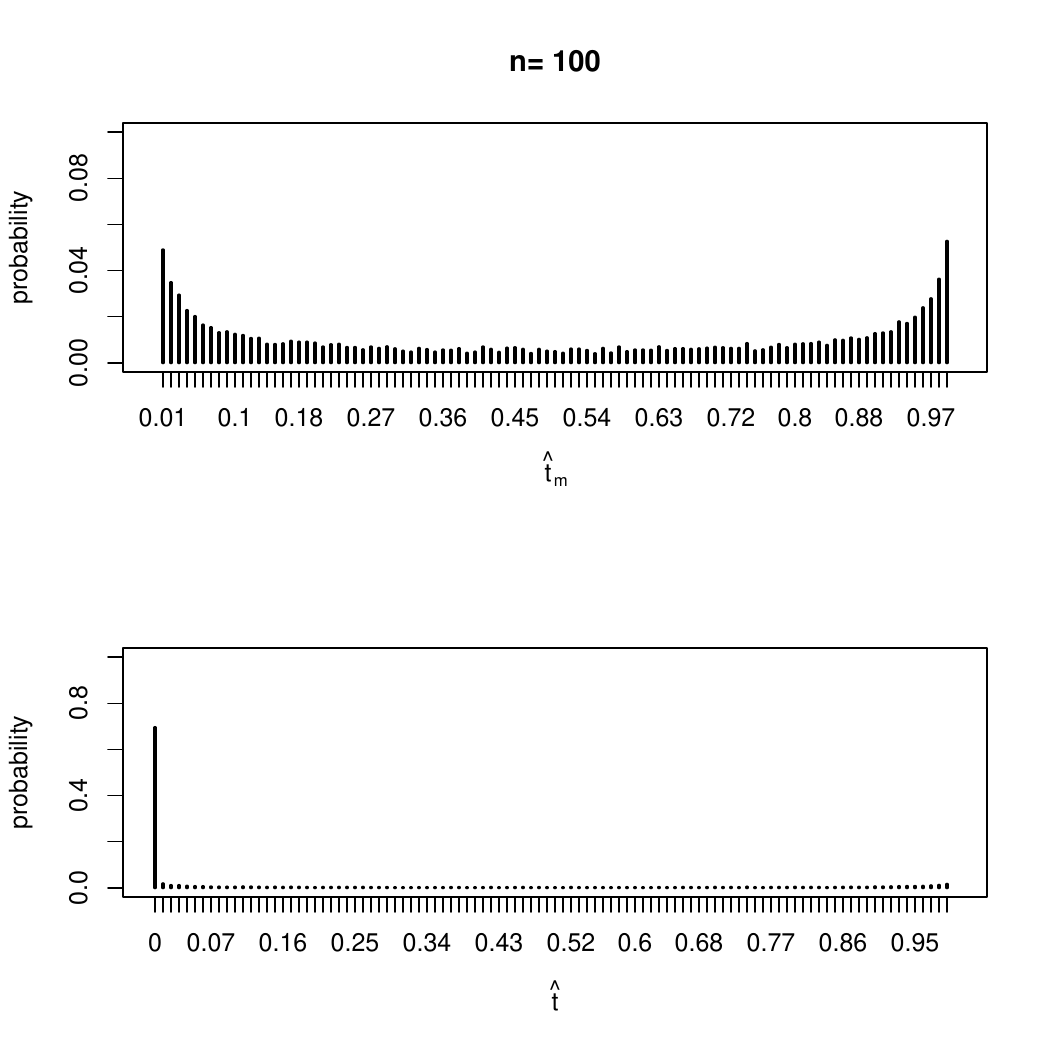}
\caption{ Distribution of $\hat{t}_m$ and $\hat t$ when sample size $n=100$ with no change-point }
\label{fig:t_hat_100}
\end{figure}

\begin{figure}[b!]
\centering
\includegraphics[height=4cm,width=5cm]{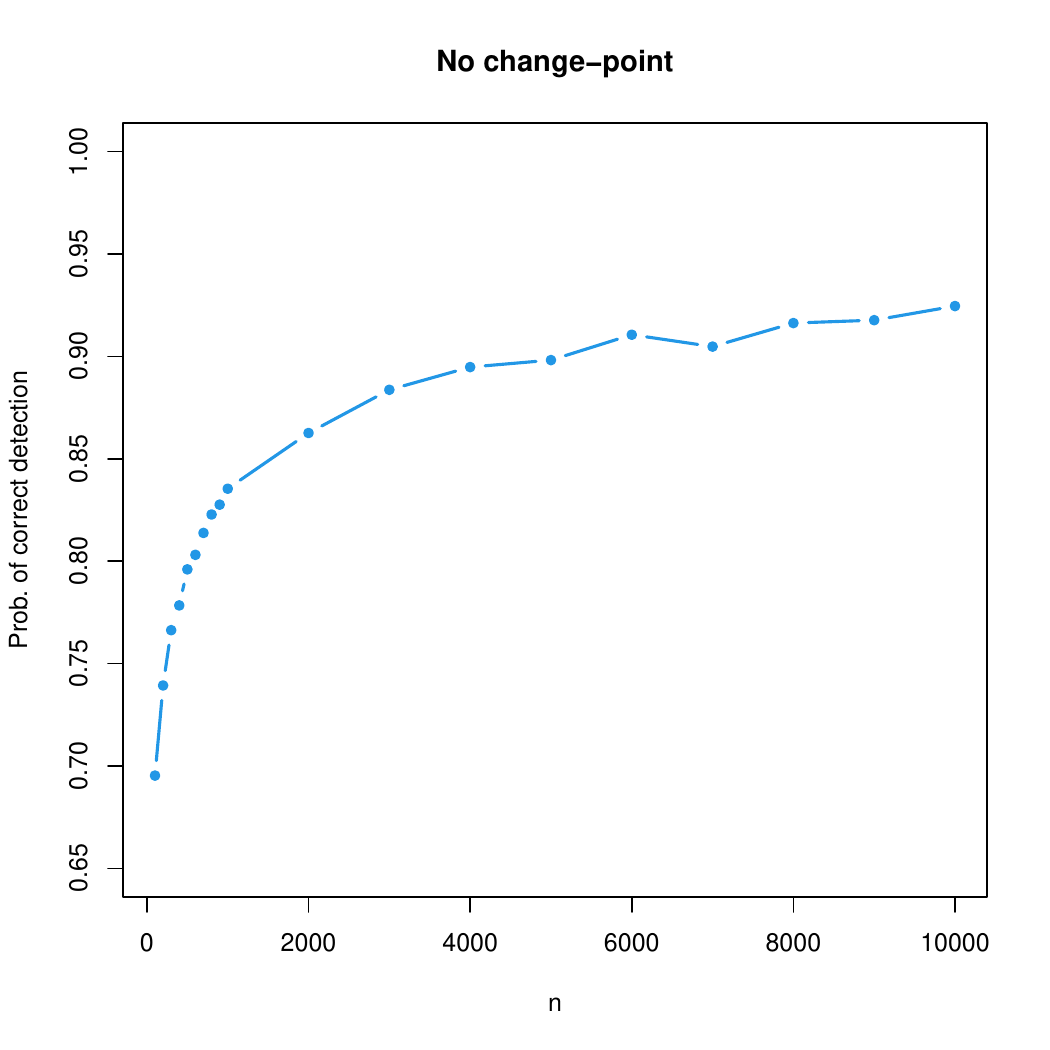}
\caption{ $P(\hat{t}=0)$ with no change-point in the data of different  sizes ($n$) from 100 to 10000 . }
\label{fig:zero_prop}
\end{figure}


\begin{figure}
\centering
\includegraphics[height=8cm,width=8cm]{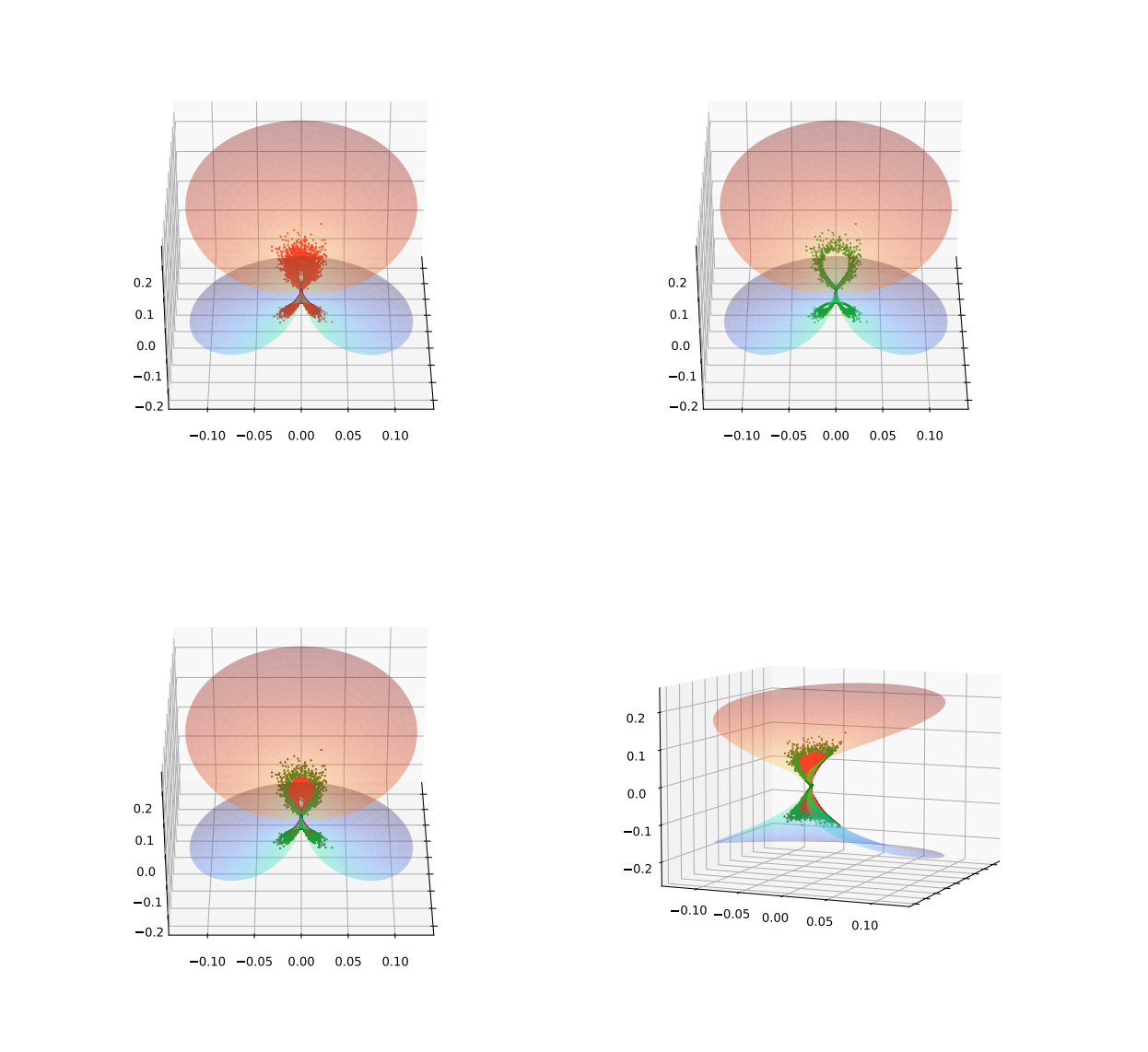}\\
\caption{ Scatter plot of $\hat {\mathbf{u}}_m$ (red, top-left) and $\hat {\mathbf{u}}$ (green, top-right)  and overlap of them on $\mathcal{M}$  in the bottom panel from two different angels when there is no change-point in data. }
\label{fig:null_dist}
\end{figure}

\subsection{Simulation Analysis}
\label{simulation}
In  Experiment (1), we  use simulation to obtain the marginal distributions of $\hat{t}_m(\equiv \frac{\hat{r}_m}{n})$ and $\hat{t}(\equiv \frac{\hat{r}}{n})$. For this purpose, we perform 10000 iterations of the following:  (a) Generate an i.i.d  sequence of 100 standard Gaussian random variates and (b) compute the estimates $\hat{t}_m$ and $\hat{t}$.  The  marginal distribution of  $\hat{t}_m$ and $\hat{t}$ are  provided in Figure \ref{fig:t_hat_100}. As expected $\hat{t}_m$ has no mass at $t=0$ whereas $\hat{t}$ has a significant mass ($\approx 70\%$) at $t=0.$ We also note that the entire mass of the distribution of $\hat{t}_m$ is spread all over the open interval $(0,1)$ in contrast to only $30\%$ of that for $\hat{t}$.  Thus the problem of MLE always giving false alarm can be mitigated to a great extent by using $\hat{t}$ instead of $\hat{t}_m$.  Figure \ref{fig:zero_prop} examines $\hat{P}(\hat{t}=0)$ i.e. the estimated probability of $P(\hat{t}=0)$  for different sample sizes varying from from 100 to 10000.

\begin{figure}[b!]
\centering
\includegraphics[height=5cm,width=5cm]{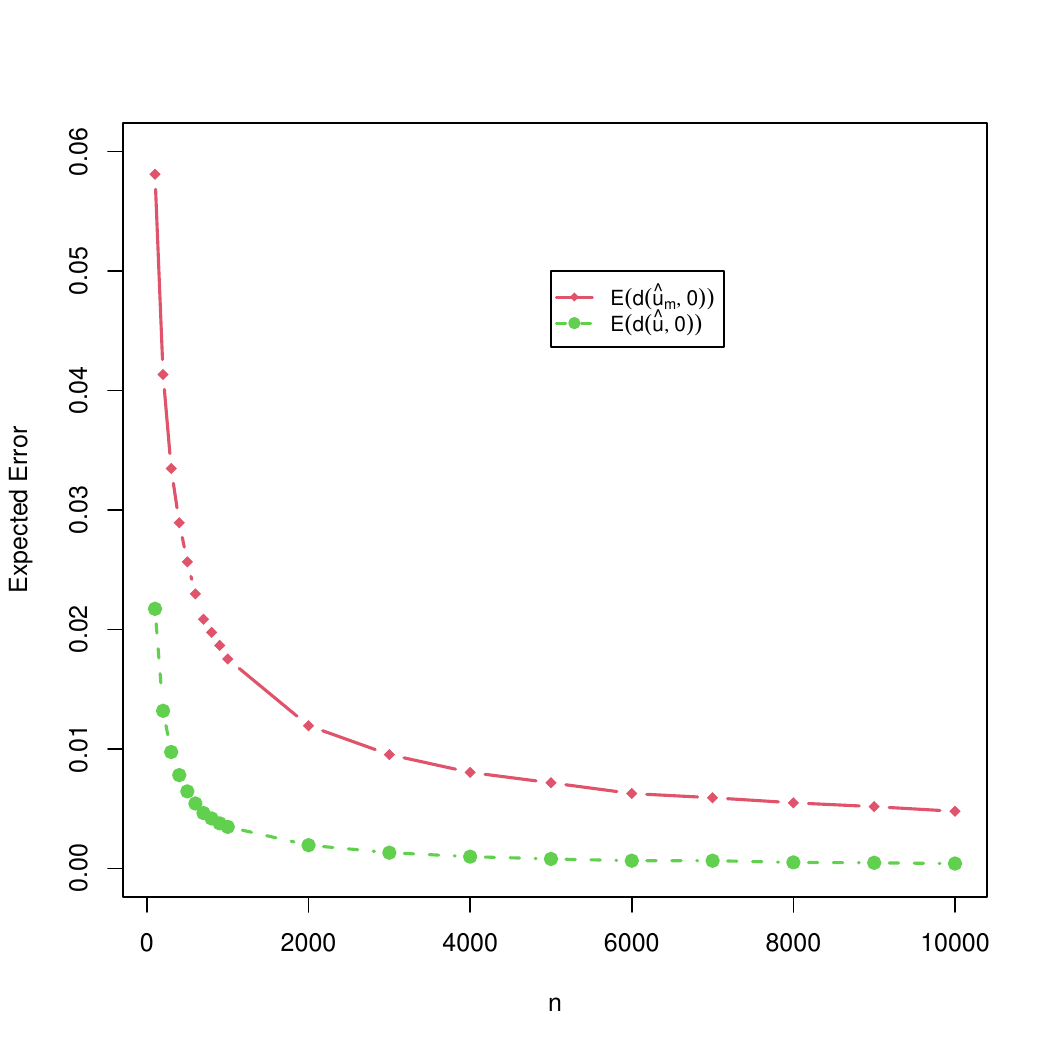}
\includegraphics[height=5cm,width=5cm]{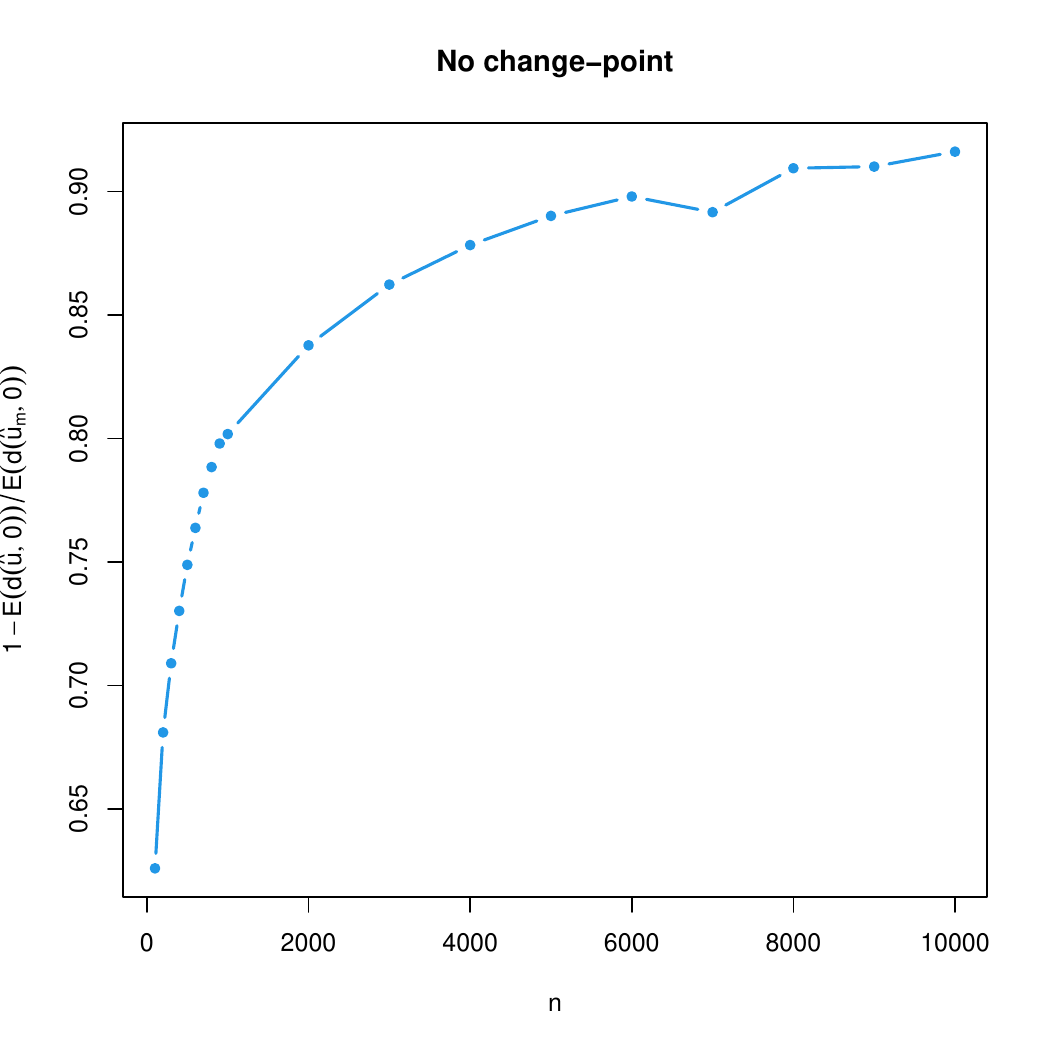}
\caption{ Expected error of the MLE $\hat{\mathbf{u}}_m$ and the proposed estimator $\hat{\mathbf{u}}$  (left) and their relative efficiency  (RE)  (right) for varying sample sizes $n$ from 100 to 100000 when there is no change-point in the data.}
\label{fig:error_nnull}
\end{figure}


\begin{figure}[b!]
\includegraphics[height=8cm,width=8cm]{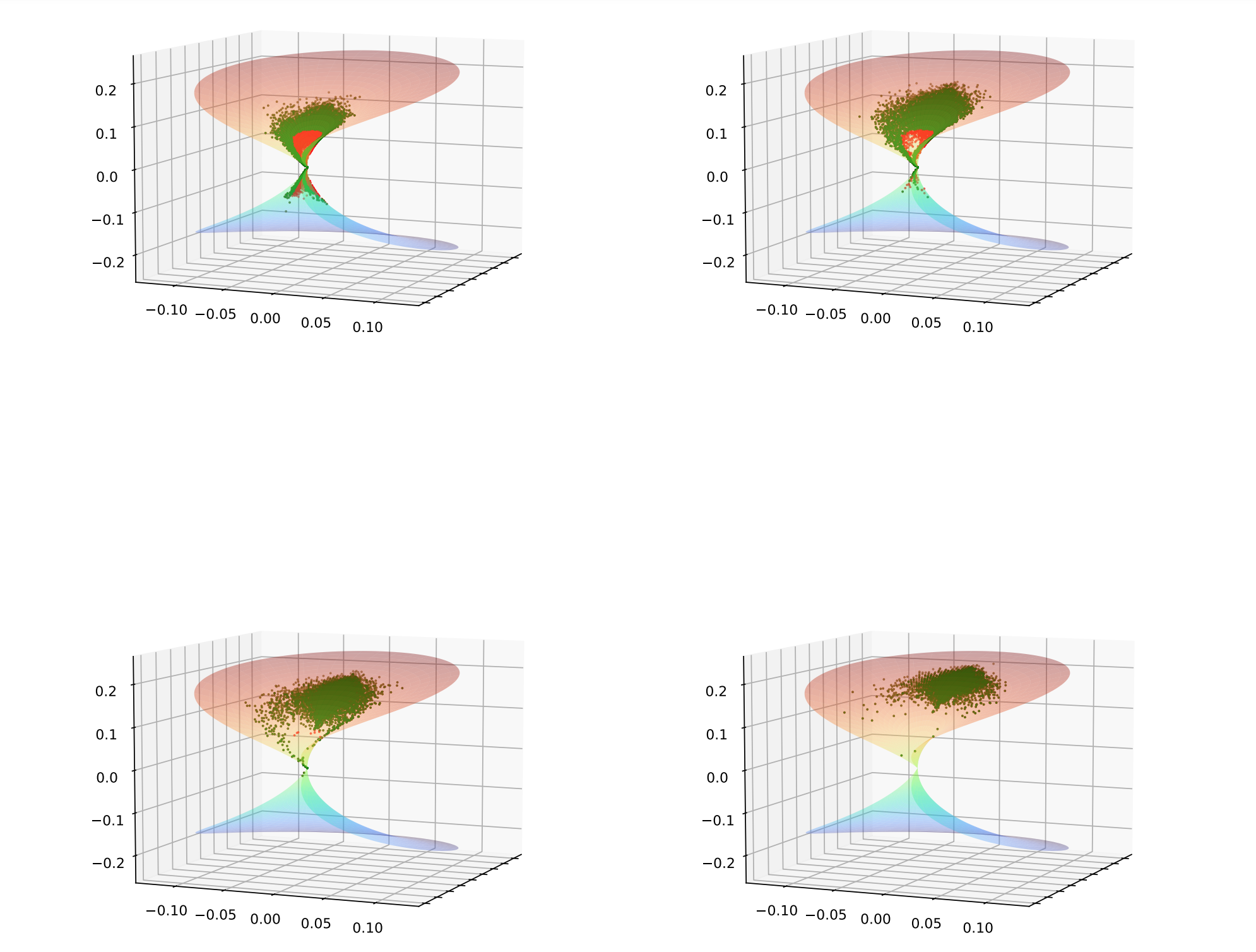}
\caption{ Scatter plot of $\hat {\mathbf{u}}_m$ (red) and $\hat {\mathbf{u}}$ (green) on $\mathcal{M}$ for sample size $n=200$ and change-point $r=100$ with $\Delta=0.25(\mbox{top-left}), 0.50 (\mbox{top-right}), 0.75 (\mbox{bottom-left}), 1.00 (\mbox{bottom-right})$ respectively.  }
\label{fig:alt_dist}
\end{figure}

In Figure \ref{fig:null_dist}  we plot  $\hat {\mathbf{u}}_m$ (red) and $\hat {\mathbf{u}}$ (green) for 10000 iterations on $\mathcal{M}$  for $n=200$ when no change point is present. As expected the top-left plot of $\hat {\mathbf{u}}_m$ shows scattering of points near $\mathbf{0}$ but excluding $\mathbf{0}$. In contrast, the top-right plot of $\hat {\mathbf{u}}$ shows a significant mass at $\mathbf{0}$. The bottom panel shows the overlap of $\hat {\mathbf{u}}_m$ and $\hat {\mathbf{u}}$ from different angles and further affirms the above observation.

It is observed from the simulation results  that the expected  \textit{zero-pass metric} error (i.e. $E(d(\cdot, \mathbf{u}))$ of  $\hat{\mathbf{u}}$  is significantly less than that of $\hat{\mathbf{u}}_m$.  We define the relative efficiency of $\hat{\mathbf{u}}$  with respect to $\hat{\mathbf{u}}_m$ as  $RE(\hat{\mathbf{u}},\hat{\mathbf{u}}_m)=1-\frac{E(d(\hat{\mathbf{u}}, \mathbf{u}))}{E(d(\hat{\mathbf{u}}_m, \mathbf{u}))}$. The top panel of Figure \ref{fig:error_nnull} provides plots of ${E(d(\hat{\mathbf{u}}, \mathbf{0}))}$ and ${E(d(\hat{\mathbf{u}}_m, \mathbf{0}))}$ when there is no change point and sample size $n$ varies  between 100 to 10000. We find that as sample size increases the expected errors of both $\hat{\mathbf{u}}$ and $\hat{\mathbf{u}}_m$ decreases with ${E(d(\hat{\mathbf{u}}, \mathbf{0}))}< {E(d(\hat{\mathbf{u}}_m, \mathbf{0}))}$ for all $n$ in the above range. From the bottom panel it is noted that relative efficiency increases with increasing $n.$


\begin{figure}[t!]
\centering
\includegraphics[height=5cm,width=5cm]{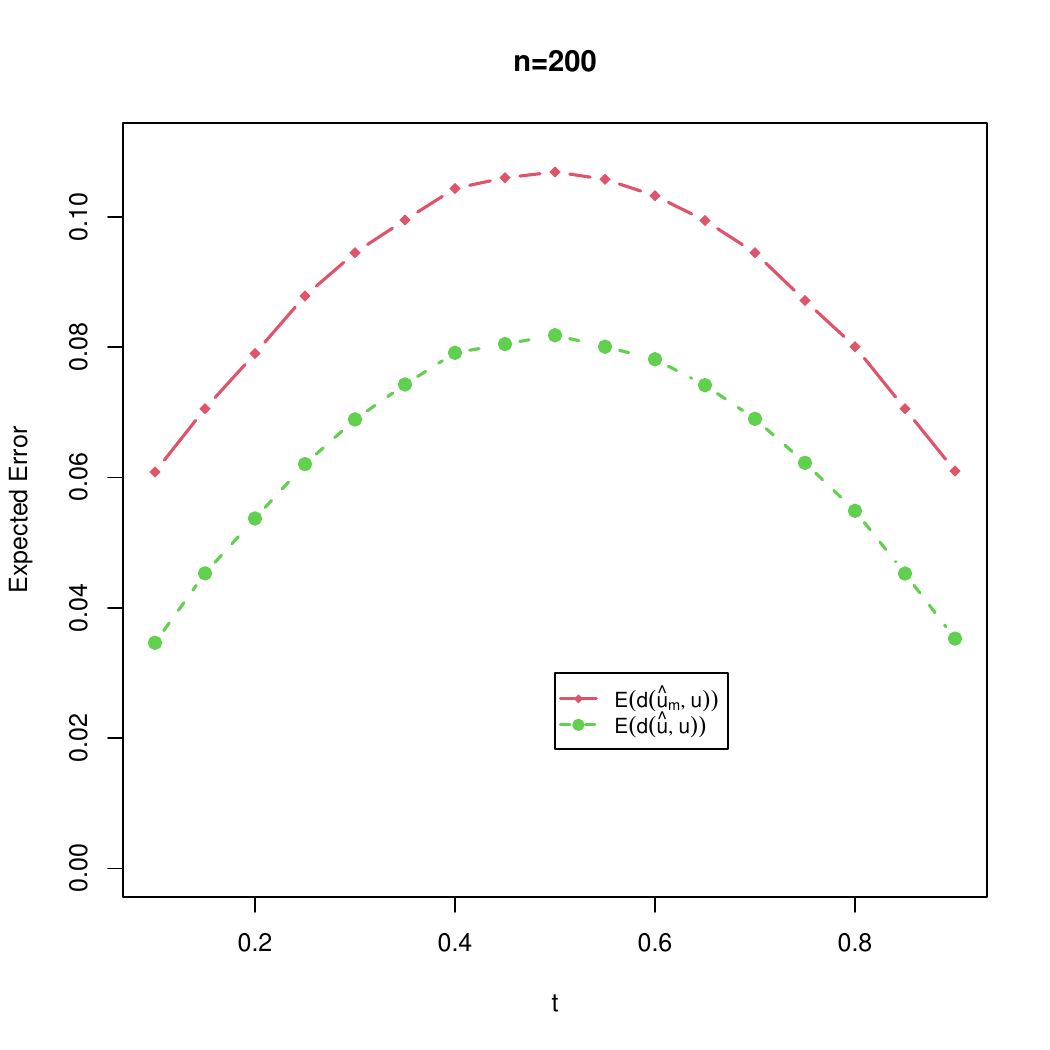}
\includegraphics[height=5cm,width=5cm]{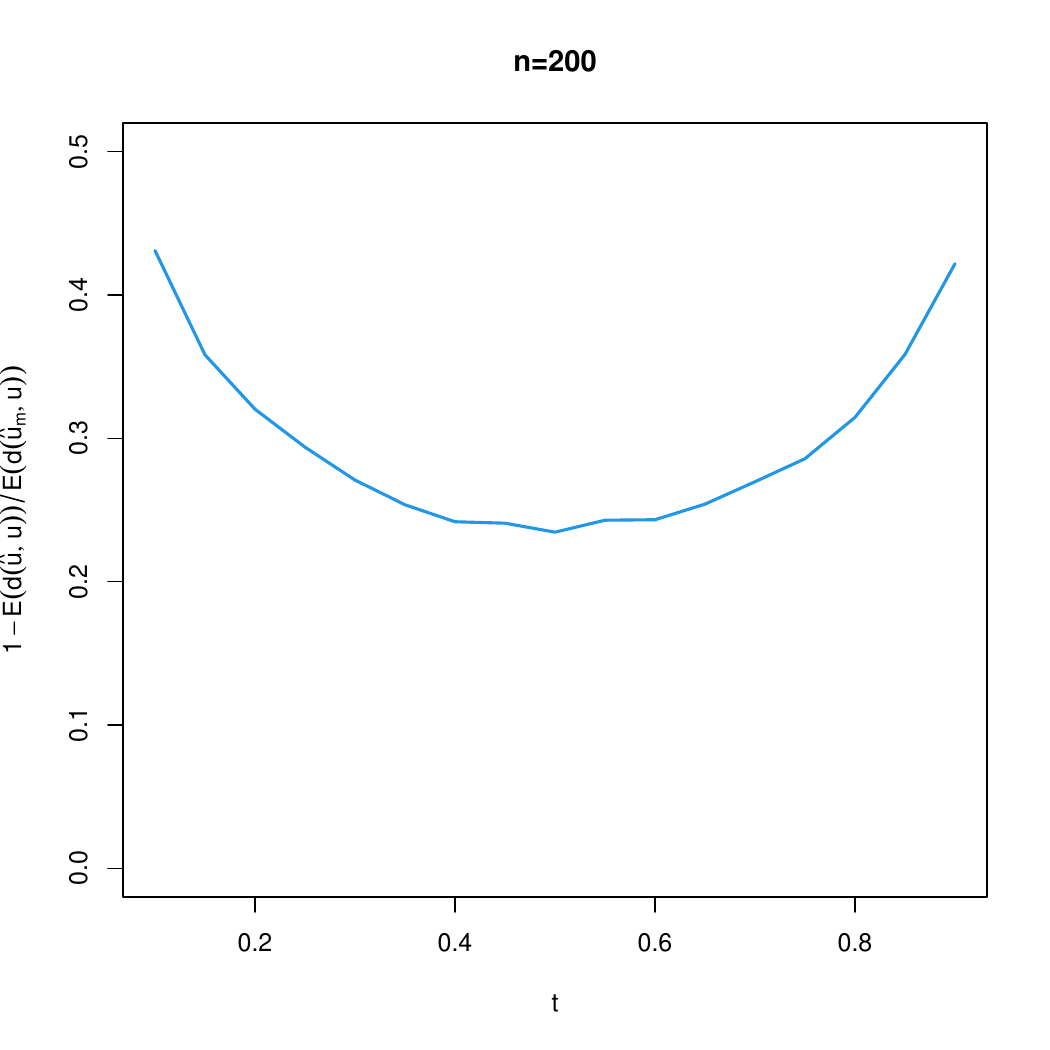}
\caption{ Expected error of the MLE $\hat{\mathbf{u}}_m$ and the proposed estimator $\hat{\mathbf{u}}$ (left) and their relative efficiency (RE) (right) for varying locations of the change-point when  $n=200$ and $\Delta=0.4$}
\label{fig:error_alt}
\end{figure}

\begin{figure}[h!]
\centering
\includegraphics[height=5cm,width=5cm]{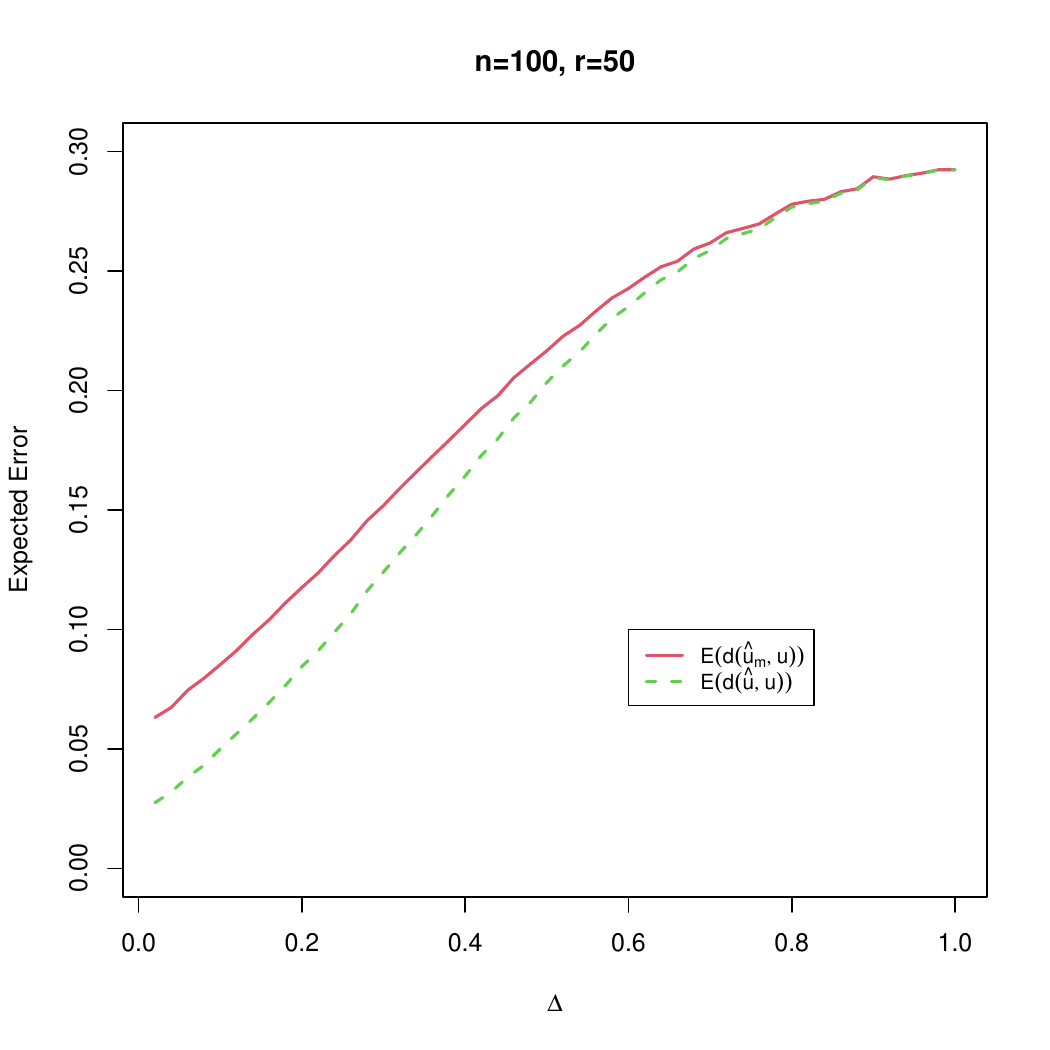}
\includegraphics[height=5cm,width=5cm]{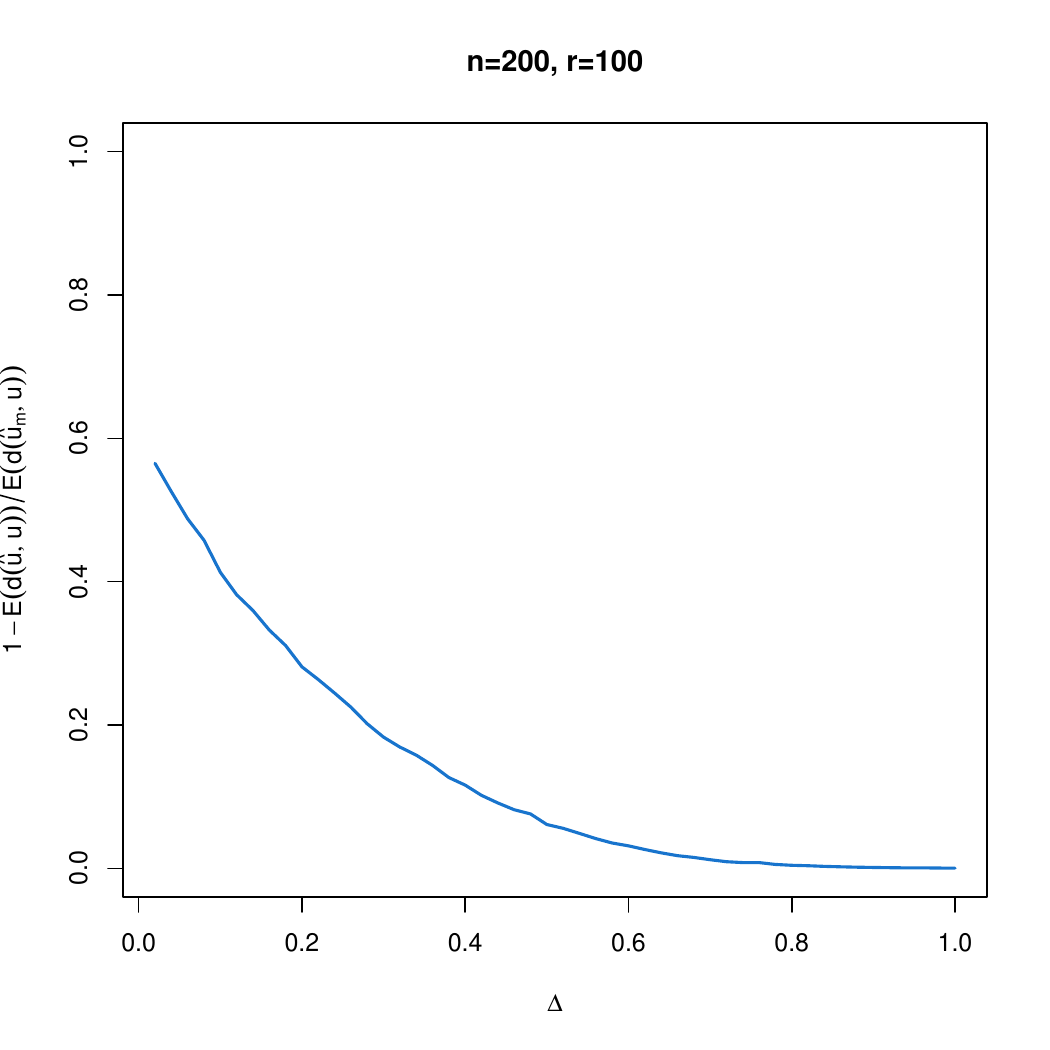}\\
\caption{ Expected error of the MLE $\hat{\mathbf{u}}_m$ and the proposed estimator $\hat{\mathbf{u}}$  (left) and their relative efficiency (RE) (right) for varying amount of mean shift $\Delta$ when sample size $n=100$ with the location of the change point at $r=50.$}
\label{fig:error100}
\end{figure}

\begin{figure}[t!]
\centering
\includegraphics[height=5cm,width=5cm]{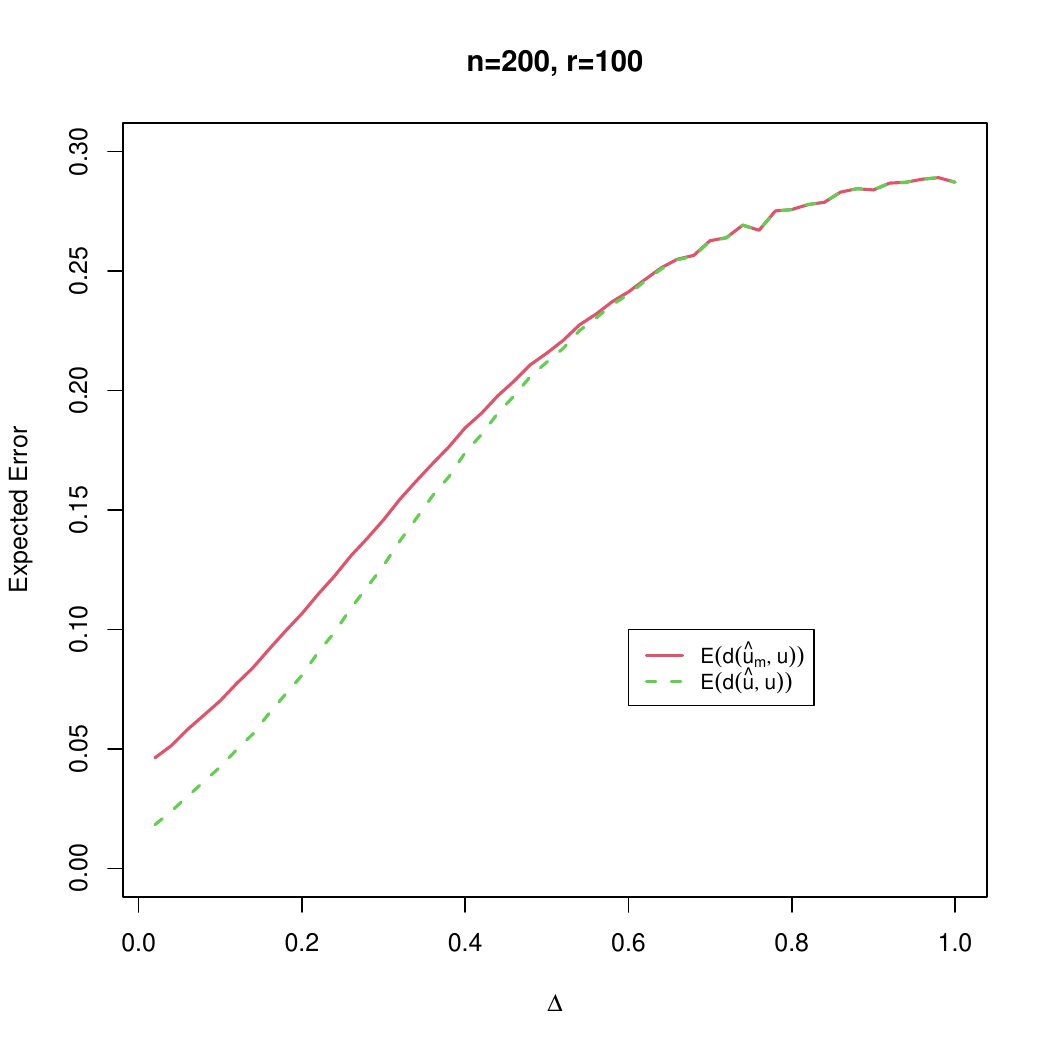}
\includegraphics[height=5cm,width=5cm]{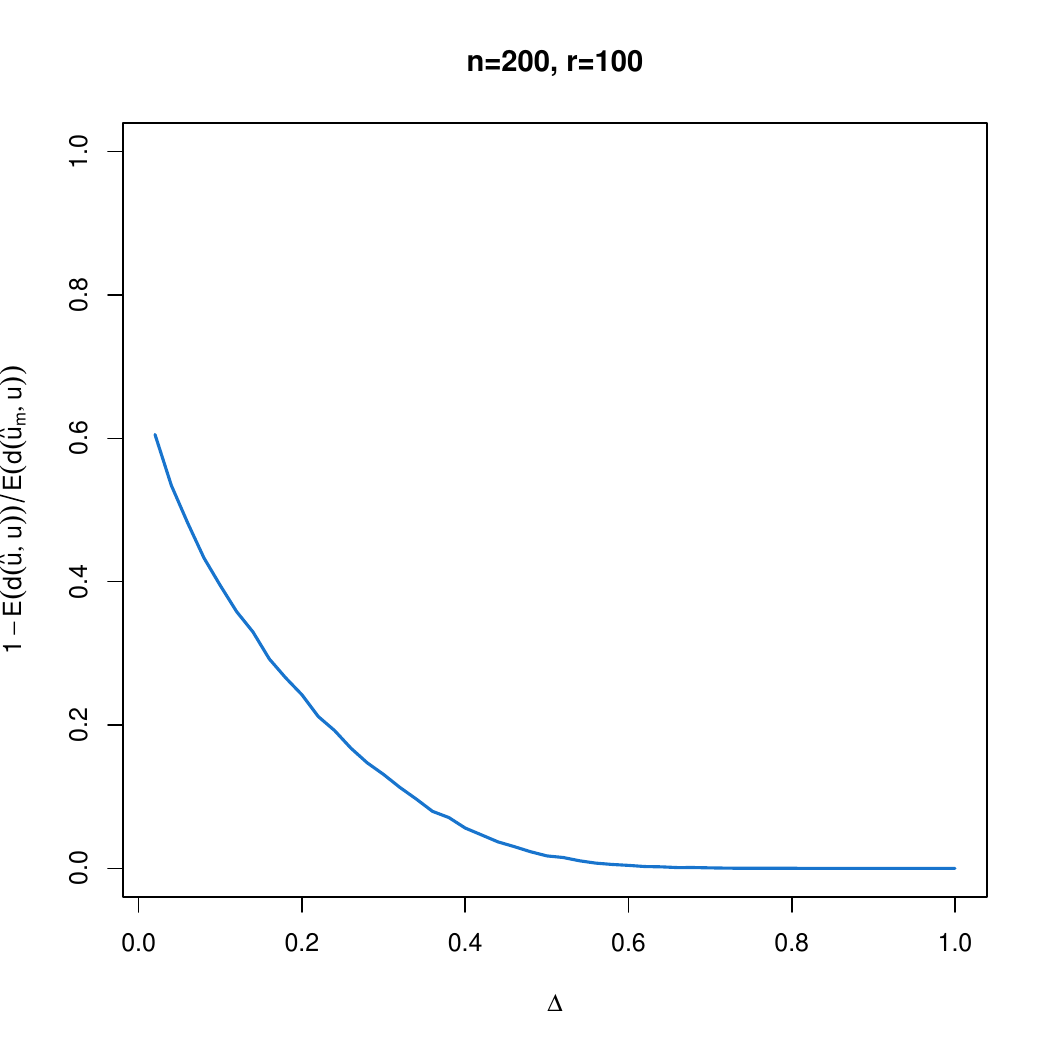}\\
\caption{ Expected error of the MLE $\hat{\mathbf{u}}_m$ and the proposed estimator $\hat{\mathbf{u}}$ (left)  and their relative efficiency (RE)  (right)for varying amount of mean shift $\Delta$ when sample size $n=200$ with the location of the change point at $r=100.$}
\label{fig:error200}
\end{figure}

In Figure  \ref{fig:alt_dist} we study the variations in the scatter plots of $\hat {\mathbf{u}}_m$ (red) and $\hat {\mathbf{u}}$ (green) on $\mathcal{M}$ with $\Delta$ for 10000 iterations and sample size $n=200$  with change-point at $r=100(\equiv(t=0.5))$.  On examining the scatter plots  $\Delta=0.25(\mbox{top-left})$, $0.50 (\mbox{top-right})$, $0.75 (\mbox{bottom-left})$ and $1.00 (\mbox{bottom-right})$ we find that the scatter plots of  $\hat {\mathbf{u}}_m$  and $\hat {\mathbf{u}}$ both move away from $\mathbf{0}$ as $\Delta$ increases. 
It is further observed from the simulation results  that  $E(d(\hat{\mathbf{u}}_m, \mathbf{u}))$ is significantly larger than  $E(d(\hat{\mathbf{u}}, \mathbf{u}))$ irrespective of the location of the change point. Figure \ref{fig:error_alt} plots $E(d(\hat{\mathbf{u}}_m, \mathbf{u}))$ and $E(d(\hat{\mathbf{u}}, \mathbf{u}))$ in the top panel  and  relative efficiency of $RE(\hat{\mathbf{u}},\hat{\mathbf{u}}_m)$ in bottom  panel for sample size $n=200$ and mean shift $\Delta=0.4$.  We note that the gain in efficiency is maximum when $t$ is close to $0$ or $1.$ 

In Figures  \ref{fig:error100} and  \ref{fig:error200}  we examine the variation of $E(d(\hat{\mathbf{u}}, \mathbf{u}))$ and $E(d(\hat{\mathbf{u}}_m, \mathbf{u}))$   with  $\Delta$ varying  from $0.1$ to $1.0$ for $n=100$ and $n=200$ respectively with change point at $\frac{n}{2}$. We find that $E(d(\hat{\mathbf{u}}, \mathbf{u}))\leq E(d(\hat{\mathbf{u}}_m, \mathbf{u}))$ for all values of $\Delta$ in this range with $\hat{\mathbf{u}}$ out-performing $\hat{\mathbf{u}}_m$ significantly for small values of $\Delta.$ The right panels in these two figures show that $RE(\hat{\mathbf{u}},\hat{\mathbf{u}}_m)$  decreases with $\Delta$ for both values of $n$. Further, it is seen that  $RE(\hat{\mathbf{u}},\hat{\mathbf{u}}_m)$ falls  faster  for $n=200$ compared to  $n=100$. 

\subsection{Data Analysis}
\label{data}
Bitcoin, introduced in 2009 by the anonymous Satoshi Nakamoto, is the first and most well-known cryptocurrency. It operates on a decentralized blockchain network, enabling secure, transparent transactions without intermediaries. With a fixed supply of 21 million coins, Bitcoin is considered a deflationary asset and a digital store of value. Its Proof of Work mechanism ensures transaction validation and network security. As a pioneer in digital finance, Bitcoin has inspired the creation of numerous cryptocurrencies and blockchain applications.

\begin{figure}[t!]
\centering
\includegraphics[height=5cm,width=10cm]{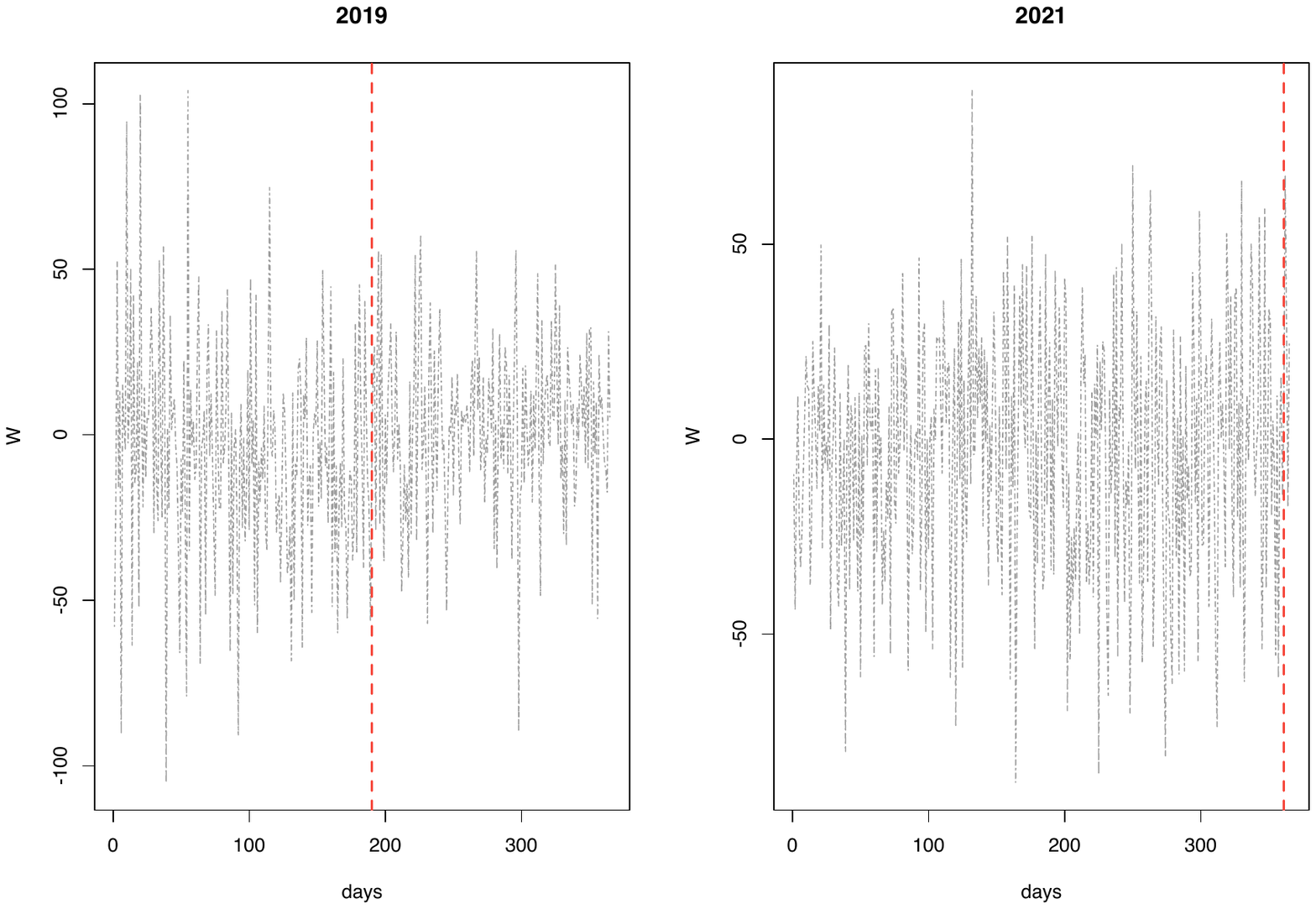}
\caption{ $W$ variable for the bitcoin data in 2019(left) and 2021(right) with  location of the change points estimated  by the MLE($\hat{r}_m$) and denoted by  vertical  red lines. For the year 2021(right) the MLE is $\hat{r}_m=361$ but  the proposed estimator shows no indication of change point. On the other hand 
in the year 2019 the  $\hat{r}=\hat{r}_m=190. $  }
\label{fig:btcdata}
\end{figure}

To evaluate our proposed change-point detection technique, we utilized a Bitcoin per-minute historical dataset sourced from Kaggle (\url{https://www.kaggle.com/datasets/prasoonkottarathil/btcinusd}). This dataset comprises one-minute historical data spanning from January 1, 2017, to March 1, 2022, resulting in 1,860 daily observations after pre-processing. The dataset includes columns for the Unix timestamp, date, symbol, opening price, highest price, lowest price, closing price, cryptocurrency volume, and base currency volume. From this data, we extracted the daily opening price at 00:00 and the closing price at 23:59. Additionally, we calculated the daily average of the lowest and highest prices over the  $24\times60$ minutes. These four variables were then used to construct a new variable for each day,
\begin{eqnarray}
    W=\frac{\mbox{Closing Price}-\mbox{Opening Price}}{\mbox{Average High Price}-\mbox{Average Low Price}}
\end{eqnarray}

It is seen that $W$ is normally distributed with the estimated variance $ (\hat{\sigma}_w^2)$ treated as known value ${\sigma}_w^2$. This is not  unrealistic in the cases discussed below as the estimate is obtained on the basis of a large sample size.   For the  year 2021  with 365 observations of $W$, estimated mean shift  is found to  have  occurred at day $\hat{r}_m=361$  as identified by the MLE. But  the proposed method does not indicate  the presence  of any  change point with  $\hat{r}={0}.$ It refers the corresponding estimates on $\mathcal{M}$  as $\hat{\mathbf{u}}_m=(2.939051\times10^{-3},  -2.026942\times10^{-4},  7.428583\times10^{-3})$  for the MLE and  $\hat{\mathbf{u}}=(0,0,0)$  for the proposed  estimator. On the contrary for the  year 2019   with 365 days the estimate by the MLE and the proposed method coincide  at $\hat{r}_m=\hat{r}=190.$ Hence the  corresponding estimated parameter $\hat{\mathbf{u}}_m=\hat{\mathbf{u}}=
(-0.010751636,  -0.001395871, 0.072761484)$ on $\mathcal{M}$. Temporal plot of the data in 2019(left) and 2021(right) are shown in Figure \ref{fig:btcdata} along with  location of the change points estimated  by the MLE($\hat{r}_m$) which is  denoted by  the vertical  red lines. 

For  the year 2019 the estimated  pooled variance $\widehat{Var(W)}=\hat{\sigma}_w^2=(29.59114)^2$ and $(\hat \mu_2-\hat \mu_1)/\hat{\sigma}_w=\hat{\Delta}=0.3079598$ are used to obtained the estimated average loss through parametric bootstrap with 10000 iterations. It is found that for the proposed estimator has the estimated risk $\widehat{E(d(\hat{\mathbf{u}}, \cdot))}=0.1381348$ and  that of the MLE is  $\widehat {E(d(\hat{\mathbf{u}}_m, \cdot ))}=0.1477256.$ Similarly for  the year 2021 the estimated  pooled variance $\widehat{Var(W)}=\hat{\sigma}_w^2=(31.22788)^2$ and $\hat{\Delta}=0.8110595$ are used to obtained the estimated risk  through  parametric bootstrap. For MLE it is observed that the estimated risk is $\widehat {E(d(\hat{\mathbf{u}}_m, \cdot ))}=0.0306724.$ On the other hand, under the consideration of no change point  $\widehat{Var(W)}=\hat{\sigma}_w^2=(32.0087)^2$  and it is found that the proposed estimator has the estimated risk $\widehat{E(d(\hat{\mathbf{u}}, \cdot))}=0.008623216.$

\section{Extended application and discussion}
\label{discussion} 
Although the proposed method has been introduced for independent observations and maximum likelihood method of estimation, the idea is extendable to the other method of estimation, e.g. CUSUM and for dependent data too for changepoint detection. If the CUSUM process is used for independent data then the change point changepoint detecting statistic is given by:
\[
\arg \max_{1 \leq k < n} H_n(k) = \arg \max_{1 \leq k < n} \left|\frac{1}{\sqrt{n}}\sum_{i=1}^{k} (X_i - \bar{X}_n)\right|,
\]
where \( \bar{X}_n = \frac{1}{n} \sum_{i=1}^{n} X_i. \)  
Now, if we replace the likelihood sequence $\mathbf{\ell}$ with $\mathbf{h}=\exp({\mathbf{h}_n})$ where  $$ \mathbf{h}_n=(H_n(0), H_n(1) , H_n(2), \ldots, H_n(n-1))$$ and follow the rest of the method as proposed in Section-\ref{estimation} we obtain the following observations. We conducted the simulation for sample size 300 and 10000 iterations. When there is no change point in the normally distributed data, then the estimator based on $H_n(\cdot)$ has the average error 0.0468, whereas the average error is negligible in the proposed method. On the other hand, when there is a change point on the data set with certain magnitude, the proposed method performs better compared to that of the CUSUM based method. The numerical comparison is in Table-\ref{tab:cusum}.

\begin{table}[h!]
\centering
\caption{Comparison of average errors  for   different location of change points\\ when $n=300$ and $H_n$ is used as an estimator}
\begin{tabular}{cccc|ccc}
\toprule
\textbf{Change point at} & \multicolumn{3}{c|}{When $\Delta=0.5 \sigma$, average error of } & \multicolumn{3}{c}{When $\Delta=0.9 \sigma$, average error of} \\
\cmidrule(r){2-4} \cmidrule(l){5-7}
& {${H_n}$} & {Proposed Method ($\mathbf{h}$) } & {Ratio} & {${H_n}$} & {Proposed Method } & {Ratio} \\
\midrule
50  & 0.1345808 & 0.0643955 & 0.4784895 & 0.1123456 & 0.0932456 & 0.5678901 \\
100 & 0.1958987 & 0.1030328 & 0.5259495 & 0.1765432 & 0.1109876 & 0.5987643 \\
150 & 0.2153522 & 0.1159119 & 0.5382435 & 0.1987321 & 0.1201234 & 0.6123456 \\
200 & 0.1950658 & 0.1030328 & 0.5281951 & 0.1776543 & 0.1098765 & 0.6054321 \\
250 & 0.1344202 & 0.0643955 & 0.4790613 & 0.1234567 & 0.0954321 & 0.5898765 \\
\bottomrule
\end{tabular}
\label{tab:cusum}
\end{table}

On the other hand when $\{X_t\}_{t=1}^n$ be a stationary time series with  finite variance  \cite{shao_2010} used the  following self-normalized statistic for identifying the   change point  in the mean at some unknown location as 
\[
\arg \max_{1 \leq k \leq n-1}G_n(k) = \arg \max_{1 \leq k \leq n-1} \frac{S_k^2}{V_n(k)/n}.
\]
where $
S_k = \sum_{i=1}^k (X_i - \bar{X}_n), \bar{X}_n = \frac{1}{n} \sum_{i=1}^n X_i.$ 
and 
for $1 \leq k \leq n-1$, the self-normalizer:
\[
V_n(k) = \sum_{t=1}^k \left( \sum_{i=1}^t (X_i - \bar{X}_k) \right)^2 + \sum_{t=k+1}^n \left( \sum_{i=k+1}^t (X_i - \tilde{X}_{n-k}) \right)^2,
\]
where $\bar{X}_k$ and $\tilde{X}_{n-k}$ denote the sample means of the first $k$ and last $n-k$ observations, respectively.
Under mild regularity conditions, it is shown when there is no change point. 
\[
G_n \xrightarrow{d} \sup_{0 < t < 1} \frac{B_0^2(t)}{ \int_0^t \left(B_0(s) - \frac{s}{t}B_0(t)\right)^2 ds + \int_t^1 \left(B_0(s) - \frac{1 - s}{1 - t} B_0(t)\right)^2 ds }=\sup_{0 < t < 1} C_0(t) \mbox{~ ~ ,say},
\]
where $B_0(t)$ is a standard Brownian bridge on $[0,1]$.
Now, if we replace the likelihood sequence $\mathbf{\ell}$ with $\mathbf{g}=\exp({\mathbf{g}_n})$ where  $$ \mathbf{g}_n=(C_0(0), C_0(1/n) , C_n(2/n), \ldots, C_0((n-1)/n))$$ and follow the rest of the method as proposed in Section-\ref{estimation}. We conducted the simulation for sample size 300 and 10000 iterations. When there is no change point in the normally distributed data, then the estimator based on $C_0(\cdot)$ has the average error 0.04255, whereas the average error is negligible in the proposed method. On the other hand, when there is a change point on the data set with certain magnitude ($\Delta= 0.25$ and $0.35$), the proposed method performs better compared to that by  \cite{shao_2010}. The numerical comparison is in Table-\ref{tab:jasa2010}. We have done the comparisons based on the limiting distribution to overcome the unavoidable influencing factor of the parameter(s) in any stationary timeseries would have been considered.

\begin{table}[h!]
\centering
\caption{Comparison of average errors  for   different location of change points\\ when $n=300$ and $C_0$ is used as an estimator}
\begin{tabular}{cccc|ccc}
\toprule
\textbf{Change point at} & \multicolumn{3}{c|}{When $\Delta=0.25$, average error of } & \multicolumn{3}{c}{When $\Delta=0.35$, average error of}  \\
\cmidrule(r){2-4} \cmidrule(l){5-7}
& {${C_0}$} & {Proposed Method ($\mathbf{g})$} & {Ratio} & {${C_0}$} & {Proposed Method } & {Ratio} \\
\midrule
50  & 0.0841811  & 0.0603974 & 0.7174702  & 0.1024093  & 0.0826626 & 0.8071788  \\
100 & 0.1162792  & 0.1010344  & 0.8688952  & 0.1501512  & 0.1417699  & 0.9441812  \\
150 & 0.1281130  & 0.1163491  & 0.9081753  & 0.1669407  & 0.1617186  & 0.9687193  \\
200 & 0.1163069  & 0.1012933  & 0.8709139  & 0.1498310  & 0.1414805  & 0.9442674  \\
250 & 0.0840086 & 0.0606829 & 0.7223413  & 0.1024924  & 0.0829970 & 0.8097868  \\
\bottomrule
\end{tabular}
\label{tab:jasa2010}
\end{table}

\section{Conclusion}
\label{conclusion}
The most widely adopted approach to the change point problem for the mean of a distribution has been that of a test for the existence of a change-point, that is followed by an estimation process of the change-point and the mean shift. Thus, in this approach,  the estimation process is inextricably intertwined with the testing process. The use of a testing procedure before embarking on the estimation problem is necessitated due to the fact that, if a straightforward maximum likelihood method is utilized, it always shows the presence of a change-point as is proved in this paper. Thus, when there is no change point in the dataset, the maximum likelihood method always raises a false alarm. Since the change-point problem is inherently an estimation problem, methods that do not involve a hypothesis test are desirable.   In this paper, we propose a new estimation approach that overcomes the shortcomings of the MLE. A unique reparametrization converts the parameter space to a horn torus with varying radius, which is a conic manifold. The location of the change point and the mean-shift are estimated using a novel random walk based estimation technique. Interestingly, the proposed estimator either indicates non-existence of a change point or coincides with the MLE.  This removes the need for an additional step of carrying out a hypothesis test.  The efficiency of the newly introduced estimation method of the mean shift is established using a Riemannian metric on the conic manifold. The proposed method is implemented on bitcoin data and its performance is compared with that of the MLE.
\newpage

\section{Appendix} 
\label{Appendix}

\subsection{Proof of the Lemma \ref{lm_mle_ratio}}
\label{pf_lm_mle_ratio}
With out loss of generality let us assume $k\in \mathbb{Z}_{n-1} \setminus \{0\}$. Let us denote 
\begin{eqnarray}
\hat\mu_n &=& arg\max_{\mu \in \Xi } \prod_{i=1}^n f(\mu|x_i)\nonumber\\
\hat\mu_{k,1} &=& arg\max_{\mu \in \Xi } \prod_{i=1}^k f(\mu|x_i)\nonumber\\
\hat\mu_{n-k,2} &=& arg\max_{\mu \in \Xi } \prod_{i=k+1}^n f(\mu|x_i)\nonumber
\end{eqnarray}

Under the assumption of unique maxima of the likelihood as stated in Lemma \ref{lm_mle_ratio}
\begin{eqnarray}
&& \max_{\mu_1,\mu_2} \prod_{i=1}^{k} f(\mu_1|x_i)\prod_{i=k+1}^n f(\mu_2|x_i) \nonumber\\
&=&\max_{\mu_1} \prod_{i=1}^{k} f(\mu_1|x_i)\max_{\mu_2} \prod_{i=k+1}^n f(\mu_2|x_i) \nonumber\\
&=& \prod_{i=1}^{k} f(\hat\mu_{k,1}|x_i) \prod_{i=k+1}^n f(\hat\mu_{n-k,2}|x_i) \nonumber\\
&>&  \prod_{i=1}^{k} f(\hat\mu_n|x_i) \prod_{i=k+1}^n f(\hat\mu_n|x_i) \nonumber\\
&=& \max_\mu \prod_{i=1}^{n} f(\mu|x_i)
\end{eqnarray}

\subsection{Proof of the Lemma  \ref{lm_cp_eatimaors}}
\label{pf_lm_cp_eatimaors}
Let $0<a,b,c$  then $a<b \Leftrightarrow a(a+c)<b(b+c)$. It is immediate that the  likelihood sequence $\boldsymbol{\ell}=(\ell(0), \ell(1), \cdots, \ell(n-1)) $, and $\mathbf{L}=\boldsymbol{\ell}/S$ have same relative order. Recall from the  Equation \ref{pi_vector} that $\pi(i)\propto L(i)(L(i)+L(0))$ for $i\neq0$. Now, in particular if we choose $a=L(i)$, $b=L(j)$ and $c=L(0)$ for arbitrary $i\neq j\in \mathbb{Z}_{n-1}\setminus \{0\}$ then $\pi(i)$ and  $\pi(j)$ preserve the same order of  $L(i)$ and  $L(j)$.  As a consequence   $\{L(1),L(2), \ldots, L(n-1)\}$ and  $\{\pi(1),\pi(2), \ldots \pi(n-1)\}$ will preserve the same relative order. But, from the Lemma \ref{lm_mle_ratio} we get  that  $L(0)< L(j)$ for all $j=1,2,\ldots, (n-1).$ As a consequence 
$$\arg\max_{k \in \mathbb{Z}_{n-1} \setminus \{0\} } \pi(k)=\arg\max_{k \in \mathbb{Z}_{n-1}  \setminus \{0\}} \ell(k)=\arg\max_{k \in \mathbb{Z}_{n-1}} \ell(k).$$
Since  
$P(\pi(0) = \max \{\pi(1),\pi(2), \ldots, \pi(n-1)\})=0$, we can consider the following two possibilities for $\pi(0)$:  
\begin{eqnarray*}
  \mbox{Case-I:~~}  \pi(0) &<& \max \{\pi(1),\pi(2), \ldots, \pi(n-1)\} \nonumber\\
  \mbox{ or Case-II:~~ }\pi(0)&>& \max \{\pi(1),\pi(2), \ldots, \pi(n-1)\} \nonumber
\end{eqnarray*}
For the Case-I we have $\hat r=\hat r_m$, and for the Case-II we have $\hat r=0$. 

\newpage
\subsection{Proof of the Theorem  \ref{thm_efficiency}}
\label{pf_thm_efficiency}

Let us define 
\begin{equation}
 \hat{\Delta}_k=
 \displaystyle	\frac{1}{n-{k}} \sum_{i={k}+1}^{n}X_i- \displaystyle	\frac{1}{k} \sum_{i=1}^{{k}}X_k \mbox{~~if~~} {k} \in \mathbb{Z}_{n-1}\setminus \{0\}
 \end{equation}
with \begin{equation}
E(\hat{\Delta}_k)=\begin{cases}
    \frac{n-r}{n-k}\Delta \mbox{\hspace{0.5cm} if ~~} k\leq r \nonumber\\
     \frac{r}{k}\Delta \mbox{\hspace{0.85cm} if ~~} k>r \nonumber
\end{cases}
\end{equation}
and $Var(\hat\Delta_k)=\frac{n}{k(n-k)}.$  On the other hand,  the log likelihood ratio can be expressed as 
\begin{equation}
-\log_e\left(\frac{\ell(0)}{\ell(k)}\right)=-\log_e\left(\frac{L(0)}{L(k)}\right)=\left(\frac{n}{2}\right)\left(\frac{k}{n}\right)\left(1-\frac{k}{n}\right)\left(\hat{\Delta}_k\right)^2. 
\label{ratio_delta}
\end{equation}
The scaled  mean difference $\sqrt{\frac{k(n-k)}{n}} \hat\Delta_k$ has variance one with expectation
\begin{equation}
E\left(\sqrt{\frac{k(n-k)}{n}}\hat{\Delta}_k\right)=\begin{cases}
    \sqrt{n\left(\frac{r}{n}\right)\left(1-\frac{r}{n}\right)\left(\frac{k}{r}\right)\left(\frac{n-r}{n-k}\right)}\Delta \mbox{\hspace{0.5cm} if ~~} k\leq r \nonumber\\
   \sqrt{n\left(\frac{r}{n}\right)\left(1-\frac{r}{n}\right)\left(\frac{r}{k}\right)\left(\frac{n-k}{n-r}\right)}\Delta\mbox{\hspace{0.85cm} if ~~} k>r. \nonumber   
\end{cases}
\end{equation}
Now considering the difference of the loss functions
\begin{eqnarray}
   & & E(d(\hat{\mathbf{u}}_m, \mathbf{u})-d(\hat{\mathbf{u}}, \mathbf{u}))\nonumber\\
    &=&\displaystyle\sum_{k=1,k\neq r}^{n-1}\left\{\left(\frac{k}{n}\right)\left(1-\frac{k}{n}\right)E|\tan^{-1}\hat{\Delta}_k| \right\}p_L(k)\delta_k\nonumber\\
    &+&\left(\frac{r}{n}\right)\left(1-\frac{r}{n}\right)\left\{E| \tan^{-1}\hat{\Delta}_r-\tan^{-1}{\Delta}|-|\tan^{-1}{\Delta}|\right\}p_L(r)\delta_r. \label{rth term}\\
     &=&\displaystyle\sum_{k=1}^{n-1}\left\{\left(\frac{k}{n}\right)\left(1-\frac{k}{n}\right)E|\tan^{-1}\hat{\Delta}_k| \right\}p_L(k)\delta_k\nonumber\\
    &+&\left(\frac{r}{n}\right)\left(1-\frac{r}{n}\right)\left\{E| \tan^{-1}\hat{\Delta}_r-\tan^{-1}{\Delta}|-(|\tan^{-1}{\Delta}|+E|\tan^{-1}\hat{\Delta}_r|)\right\}p_L(r)\delta_r.\nonumber \\ 
    &\equiv&\displaystyle\sum_{k=1}^{n-1}\left\{\left(\frac{k}{n}\right)\left(1-\frac{k}{n}\right)E|\tan^{-1}\hat{\Delta}_k| \right\}p_L(k)\delta_k +\eta_{n}(r)   
    \label{rth term diff}
\end{eqnarray}
If $\mathbf{u}=\mathbf{0}$ i.e. $r=0$ or $\Delta=0$ then  the second term of Equation \ref{rth term} is non negative. 
Now to understand the contribution of the second term of Equation \ref{rth term}  when $\mathbf{u}\neq \mathbf{0}$ we focus on the  conditional probability  
\begin{eqnarray}
    \delta_r&=&P(\hat{r}=0|\hat{r}_m=r)\nonumber\\
            &=& P(\pi(0)>\pi(r))\nonumber\\
            &=& P(L(0)>L(r)(L(r)+L(0))) \nonumber\\
            &=&P\left(\left(1+\frac{\ell(r)}{\ell(0)}\right)^{-1}>\frac{\ell(r)}{S}\right)\nonumber\\
            &=& P\left(\frac{1}{1+\exp\left\{{0.5}\left(\sqrt{\frac{r(n-r)}{n}}\hat{\Delta}_r\right)^2\right \}}> \frac{\ell(r)}{S}\right) \nonumber\\
            &\leq&P\left(\frac{1}{1+\exp\left\{{0.5}\left(\sqrt{\frac{r(n-r)}{n}}\hat\Delta_r\right)^2\right \}}
            \geq \frac{1}{n}\right) \mbox{~since~} \frac{\ell(r)}{S}= \frac{\displaystyle\max_k\ell(k)}{\displaystyle\sum_{k=0}^{n-1}\ell(k)} \geq \frac{1}{n}\nonumber\\
            &=&P\left(\hat{\Delta}^2_r\leq \frac{2\log_e(n-1)}{n} \left\{\left(\frac{r}{n}\right)\left(1-\frac{r}{n}\right)\right\}^{-1} \right) \label{pro_rth_term}
\end{eqnarray}
The right hand side of the inequality in \ref{pro_rth_term} goes to zero as $n\rightarrow \infty$ as 
$$ \frac{2\log_e(n-1)}{n} \rightarrow 0 \mbox {~and~} \left\{\left(\frac{r}{n}\right)\left(1-\frac{r}{n}\right)\right\} \rightarrow \nu(1-\nu)$$

Note that $\hat{\Delta}_r^2$ converges to $\Delta^2(\neq 0)$ in probability. As a result,  the probability in \ref{pro_rth_term} converges to zero, implying $\delta_r\rightarrow 0$ as $n\rightarrow \infty$. 
Since $\hat\Delta_r \rightarrow \Delta$ in probability  and $h_1(x)=|\tan^{-1}(x)-\tan^{-1}\Delta|$ and $h_2(x)=|\tan^{-1}(x)|$ both are  bounded continuous functions, hence by Theorem 5 on page 79 of \cite{chandra1999first} $E| \tan^{-1}\hat{\Delta}_r-\tan^{-1}{\Delta}|\rightarrow 0$ and $E|\tan^{-1}\hat{\Delta}_r|\rightarrow E|\tan^{-1}{\Delta}|.$
Thus,
$$\left(\frac{r}{n}\right)\left(1-\frac{r}{n}\right)\left\{E| \tan^{-1}\hat{\Delta}_r-\tan^{-1}{\Delta}|-(|\tan^{-1}{\Delta}|+E|\tan^{-1}\hat{\Delta}_r|)\right\}\rightarrow -2\nu(1-\nu) |\tan^{-1}{\Delta}| 
$$
Thus, the second term in the Equation \ref{rth term} goes to zero as  $n\rightarrow 
\infty$. Now observe that the first term of \ref{rth term} can be considered as 

\begin{eqnarray}
     &&P(\hat{r}=0)\displaystyle\sum_{k=1}^{n-1}\left\{\left(\frac{k}{n}\right)\left(1-\frac{k}{n}\right)E|\tan^{-1}\hat{\Delta}_k| \right\}\frac{p_L(k)\delta_k}{P(\hat{r}=0)} \mbox{~~~when~} P(\hat{r}=0)\neq0\nonumber\\
    &=&P(\hat{r}=0) E_{\hat{r}_m|\hat{r}=0}\left\{\left(\frac{{\hat{r}_m}}{n}\right)\left(1-\frac{{\hat{r}_m}}{n}\right)|\tan^{-1}\hat{\Delta}_{\hat{r}_m}| \right\} \mbox{~ using the Lemma \ref{lm_r_zero}} \nonumber\\
    &\geq &  \displaystyle \left\{\left(\frac{M}{n}\right)\left(1-\frac{M}{n}\right)E\left(\tan^{-1}|\hat{\Delta}_M|\right)  \right\}{p_L(M)\delta_M} \mbox{~~~where~} M=\arg\max_{k} \left\{p_L(k)\delta_k \right\}\nonumber\\   
\end{eqnarray}

Thus, $E(d(\hat{\mathbf{u}}_m, \mathbf{u})-d(\hat{\mathbf{u}}, \mathbf{u})) \ge \eta_n(r)$ where $\eta_n(r) \rightarrow 0$ as $n \rightarrow \infty$.

\begin{eqnarray}
    &&\displaystyle\sum_{k=1}^{n-1}\left\{\left(\frac{k}{n}\right)\left(1-\frac{k}{n}\right)E|\tan^{-1}\hat{\Delta}_k| \right\}p_L(k)\delta_k\nonumber\\
     &=&P(\hat{r}=0) E_{\hat{r}_m|\hat{r}=0}\left\{\left(\frac{{\hat{r}_m}}{n}\right)\left(1-\frac{{\hat{r}_m}}{n}\right)|\tan^{-1}\hat{\Delta}_{\hat{r}_m}| \right\} \nonumber\\
    &=&P(\hat{r}=0)\displaystyle\sum_{k=1}^{n-1}\left\{\left(\frac{k}{n}\right)\left(1-\frac{k}{n}\right)E\left(\tan^{-1}|\hat{\Delta}_k|\right)  \right\}\frac{p_L(k)\delta_k}{P(\hat{r}=0)} \mbox{~~~when~} P(\hat{r}=0)\neq0\nonumber\\
    &\geq &  \displaystyle \left\{\left(\frac{M}{n}\right)\left(1-\frac{M}{n}\right)E\left(\tan^{-1}|\hat{\Delta}_M|\right)  \right\}{p_L(M)\delta_M} \mbox{~~~where~} M=\arg\max_{k} \left\{p_L(k)\delta_k \right\}\nonumber\\
\end{eqnarray}



\section*{Acknowledgments}
Author Buddhananda Banerjee would like to thank the Science and Engineering Research Board (SERB), Department of Science \& Technology, Government of India, for the MATRICS grant (File number MTR/2021/000397)  for the project's funding. Both the authors would like to thank Mr. Surojit Biwas for collecting and processing the data to make it useful for this article. 



\newpage
\bibliographystyle{abbrvnat}
\bibliography{buddha_bib}

\end{document}